\documentclass[aps,prd,nofootinbib,10pt]{revtex4-1}
\usepackage[colorlinks=true,linkcolor=blue,citecolor=blue,urlcolor=blue]{hyperref}
\usepackage{natbib}
\usepackage[T1]{fontenc}
\usepackage{graphicx}
\usepackage{subcaption}
\usepackage{empheq}
\usepackage{slashed}
\usepackage{amsmath,float}
\usepackage{amsfonts}
\usepackage{amssymb}
\usepackage{xcolor}
\usepackage{framed}
\usepackage[toc,page]{appendix}
\usepackage{braket}
\usepackage{caption}
\usepackage{color}
\usepackage[normalem]{ulem}

\def\beq{\begin{equation}}
\def\eeq{\end{equation}}
\def\barr{\begin{eqnarray}}
\def\earr{\end{eqnarray}}

\begin{document}
\title{Neutrino Floor in Leptophilic $U(1)$ Models: Modification in $U(1)_{L_{\mu}-L_{\tau}}$ }

\author{Soumya Sadhukhan}
\email{physicsoumya@gmail.com}
\affiliation{Ramakrishna Mission Residential College (Autonomous), Narendrapur, Kolkata 700103, India} 
\affiliation{Department of Physics and Astrophysics, University of Delhi, Delhi 110 007, India}

\author{Manvinder Pal Singh}
\email{manvinderpal666@yahoo.com}
\affiliation{Department of Physics and Astrophysics, University of Delhi, Delhi 110 007, India}

\begin{abstract}
In this work, we investigate the beyond standard model (BSM) impact of leptophilic U(1) models, namely $ U(1)_{L_\mu-L_e}$, $U(1)_{L_e-L_\tau}$ and $U(1)_{L_\mu-L_\tau}$ on coherent elastic neutrino-nucleus scattering (CE$\nu$NS) and hence its effect on dark matter (DM) direct detection experiments. 
Imposing the latest relevant experimental constraints on these models, we obtain $\mathcal{O}(50\%)$ enhancement for case of $U(1)_{L_\mu-L_\tau}$ in a region $m_Z' \approx 20~$MeV. Subsequently, we observe that the enhancement seen in CE$\nu$NS is roughly getting translated to enhancement by a factor of 2.7 (for Germanium based detectors) and 1.8 (for Xenon based detectors) in the neutrino scattering event rate which eventually enhances the neutrino floor by same amount. 
This enhancement is more prominent in the region with DM masses less than 10 GeV. The model parameter space that leads to this enhancement, can simultaneously explain both anomalous magnetic moment of muon ($(g-2)_{\mu}$) and observed DM relic density, in a modified scenario. Enhancement of neutrino floor requires increased number of DM-nucleon scattering events in the future DM direct detection experiments, to establish themselves to be DM signal events. In absence of any DM signal, those experiments can directly be used to measure the neutrino rate, quantifying the BSM effects.
\end{abstract}

\maketitle

\section{Introduction and Motivation}

The majority of the matter present in our Universe is in the form of a non-luminous matter called dark matter (DM). Its presence is well motivated through astro-physical observations like galactic rotational curves and gravitational lensing etc. Particle candidates of DM are well motivated by WIMP (weakly interacting massive particle) miracle, where we expected a DM at TeV scale with interaction strength typical to have correct DM relic density. Such DM candidates were incorporated in beyond the standard model (BSM) theories in numerous ways; Inert Higgs Doublet, Right handed neutrino and Super-symmetry are few of them to be named. But till date, no conclusive observational evidence of the presence of such a particle is found either in the LHC, specifically designed to probe the TeV scale physics, or in the DM direct and indirect detection experiments. The search for the DM particles are on through different DM direct detection experiments, albeit with a renewed vigor directed to find DM particles at a lower mass scale. 

 DM direct detection relies on the measurement of its recoil energies due to the DM scattering with detector material. While direct detection experiments like Xenon1T\cite{Aprile:2015uzo}, PandaX\cite{Cui:2017nnn}, LUX \cite{Akerib:2016vxi} etc are yet to find evidence of the DM, one silicon based CDMS-II \cite{Agnese:2013rvf} detector reported three dark matter scattering events, which are in conflict with null observation from other experiments. The possibility of these three events coming from fluctuation of the standard background to the DM signal is significantly low $(\sim 5.4 \% )$~\cite{Agnese:2013rvf}. However, if presence of beyond the SM physics can substantially modify the known background then a stronger argument can be made in favor of null results from other direct detection experiments.    
 
Being neutral and weakly interacting, similar to how the DM candidates also interact, neutrino recoil can mimic the DM signal. Therefore, the neutrino events can pose as significant background to the DM events, aided by their relative prevalence in the nature, i.e. the high flux rate of the solar neutrinos. Direct detection experiments involve signals with nuclear recoil energies upto 100 keV. With this scale of nuclear recoil, the momentum transfer is sufficiently small so that scattering amplitudes from individual nucleons can coherently add up to provide the $\nu$-nuclear scattering, enhanced proportionally to total number of nucleons. This type of scattering, as observed in recent COHERENT\cite{Akimov:2017ade} experiment, is known as coherent elastic neutrino nucleus scattering (CE$\nu$NS). With increasing sensitivity and exposure of the direct detection experiments, DM exclusion plots are excluding more of the parameter space and approaching the parameter region where it will become difficult to differentiate (with 90 \% C.L.) DM-nucleus scattering events from the neutrino-nucleus ones. This region in the $\sigma_n^0 - m_{DM}$ plane, where the neutrino background remains indistinguishable from possible DM signals is termed {\it Neutrino Floor}. Any significant enhancement in the neutrino floor can raise the background in DM direct detection experiments and can therefore lead to fake positive DM signal events. Even in the absence of DM signal detection, DM experiments still can be used to directly probe the different neutrino flux induced events, once the experiments become sensitive to neutrino floor deciphering the profile. Digging deep into the floor profile can shed some light on Non Standard Interactions (NSI) in the neutrino sector.

In this work, we investigate a set of leptophilic models, $U(1)_{L_\mu-L_e}$, $U(1)_{L_e-L_\tau}$  and $U(1)_{L_\mu-L_\tau}$, where neutrino quark couplings arise only at one loop level and, due to this suppression, are therefore expected to modify the neutrino-nucleus recoil rate and the neutrino floor minimally. Still, a provision of a very light $Z^{\prime}$ is still there in these models, as this suppression can lead to relaxed constraints from proton beam dump and hadronic colliders. Further, due to absence of $Z'$ boson couplings to the $e^+/e^-$ in the case of $U(1)_{L_\mu- L_\tau}$ case, electron beam dump experiments put no constraints on low $Z^{\prime}$ mass region of the parameter space, where that is ruled out for other $U(1)$ models listed above. $U(1)_{L_\mu- L_\tau}$ model can explain the anomalous magnetic moment of muon, i.e. $(g-2)_{\mu}$~\cite{Bauer:2018onh} in the sub-GeV $m_{Z^{\prime}}$ parameter space, which is also central to presence of a light DM with observed relic density. These $U(1)$ models are also well motivated by results from DM indirect detection experiments (i.g.  DAMPE \cite{TheDAMPE:2017dtc}  and AMS02 \cite{PhysRevLett.110.141102} etc ), along with the possible explanation of $e^+/e^-$ excess observed in cosmic rays through DM annihilation to leptons via $Z^{\prime}$.
In this parameter region of $U(1)_{L_\mu- L_\tau}$, there is an extra contribution through $Z^{\prime}- \gamma$ in the CE$\nu$NS process, paving way to its significant enhancement compared to the SM value, aided by the lightness of the $Z^{\prime}$ boson. This can potentially lead to excess amount of neutrino recoils, preferably in the low recoil energy domain. This increment can essentially lead to an enhancement in the neutrino events background present in DM direct detection experiments, which translates to more DM-nucleon cross section region not being  viable to distinguish DM events from neutrino events, therefore, resulting in an upliftment of the neutrino floor. 
 
Any new interactions that can modify CE$\nu$NS can also potentially alter the neutrino floor profile. Effective operators inducing Non Standard Interactions (NSI) \cite{Chao:2019pyh,Heeck:2018nzc} between  neutrinos  and electron/quarks have been studied in connection to this. It was observed vector and scalar current NSIs show significant enhancements, especially for scalar case with augmentation of $\mathcal{O}$(20\%)  in the neutrino floor. Simplified models involving new mediators also have been studied in this context \cite{Boehm:2018sux,Bertuzzo:2017tuf}. Amplification in neutrino floor by several orders in case of scalar mediator and by a factor of two in case of vector mediator were seen with DM mass less than 10 GeV. Studies also exist where $Z'$ boson arising in U(1)$_{X}$ models such as B-L and B-L$_{(3)}$\cite{Boehm:2018sux},  can induce direct tree level coupling between neutrinos and quarks that can modify the neutrino floor. 
 
Plan of the paper is as follows. In the section~\ref{sectionmodel}, we briefly discuss the model details and Lagrangian interaction terms of $U(1)_{L_\mu-L_e}$ ,$U(1)_{L_e-L_\tau}$  and $U(1)_{L_\mu-L_\tau}$ models. 
Constraints on the parameter space in these models are also briefly discussed. 
In section~\ref{nuNucleusIR}, we investigate the modification of CE$\nu$NS rate along with combined experimental constraints in the models. Next we study the change in CE$\nu$NS event rate induced by incoming neutrino flux for the case $U(1)_{L_\mu-L_\tau}$ compared to the SM, which can appear as a background to DM signal in direct detection experiments. We choose Germanium and Xenon based detectors for their ability to scan different parameter regions of dark matter mass. In the section~\ref{sectionneutrinoFloor}, we study the modification of neutrino floor and investigate its impact in future dark matter experiments. Finally, in section~\ref{Summary and Conclusion} we summarize, along with a discussion of results.


\section{Model}
\label{sectionmodel}
In this article we have considered the minimal U(1)$_{X}$ extensions to standard model which could lead to significant non standard interaction between neutrino and nucleus which can serve as background to direct detection of dark matter.  Minimal standard model with three generation gives rise to four independent global U(1) symmetries, electron-lepton number (U(1)$_{L_e}$), muon-lepton number (U(1)$_{L_\mu}$), tau-lepton number (U(1)$_{L_\tau}$) and baryon number U(1)$_{B}$, out of which, three combinations, namely U(1)$_{L_\mu-L_e}$ ,U(1)$_{L_e-L_\tau}$  and U(1)$_{L_\mu-L\tau}$ are  free of gauge anomaly \citep{Foot:1990mn,He:1990pn,He:1991qd,Choudhury:2020cpm} without extending the SM with extra particles. U(1)$_{B-L}$ is also anomaly free with introduction of right-handed neutrinos \citep{Bauer:2018onh} and leads to modification of Neutrino floor, but we will refrain from discussing it here as it has already been discussed in \cite{Boehm:2018sux}.  In what follows we will denote U(1)$_{L_i-L_j}\equiv $  U(1)$_{i-j}$ models such that U(1)$_{L_\mu-L_\tau}\equiv$  U(1)$_{\mu-\tau}$  .

 In U(1)$_{i-j}$ models , the additional U(1) symmetry can be spontaneously broken by introduction of a new scalar $S$, which leads to the $Z'$ boson obtaining a finite mass via a non-trivial coupling to $S$ \cite{Bi:2009uj}. With these new particles we can write the additional terms besides the SM as,

 \begin{eqnarray}
L_{new}=&& -\frac{1}{4} Z'^{\mu\nu}Z'_{\mu\nu} + \sum_l \bar{l}\gamma^\mu\left(-g_{i-j}\,Y'_l\, Z'_\mu\right)l \nonumber
\\
&&+ \left(D_\mu S\right)^\dagger\left(D^\mu S\right) + \mu^2_S S^\dagger S + \lambda_{S}\left(S^\dagger S\right)^2 + \lambda_{SH}\left(S^\dagger S\right)H^\dagger H 
\end{eqnarray}
here $S$ is the new scalar where $\mu^2_S$ and $\lambda_S$ are co-efficients of bilinear and quartic self interactions respectively, which couples with SM Higgs $H$ via quartic coupling $\lambda_{SH}$ . $Z'$ boson couples with leptons $l$ through Y$_l'= L_i-Lj$ for respective U(1)$_{i-j}$ model, highlighted by the interaction term,
\begin{eqnarray}
\label{intLij}
L_i - L_f = - g_{i-j}(  \bar l_i \gamma^\mu l_i - \bar l_j \gamma^\mu l_j
+ \bar \nu_i \gamma^\mu L \nu_i - \bar \nu_j \gamma^\mu L \nu_j) Z^\prime_\mu\;.
\end{eqnarray}

\begin{figure}[h!]
\centering 
\begin{subfigure}{.49\textwidth}\centering
  \includegraphics[width=\columnwidth]{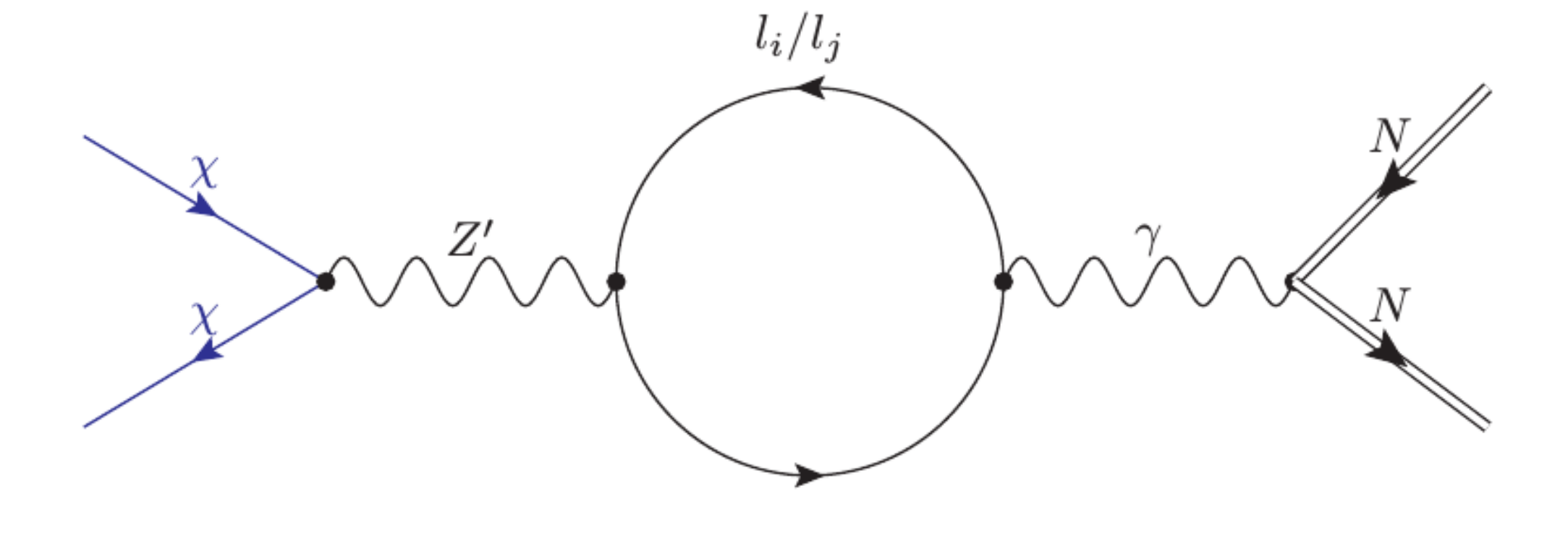}
\caption{}\label{feyn2}
\end{subfigure}%
\begin{subfigure}{.49\textwidth}\centering
  \includegraphics[width=\columnwidth]{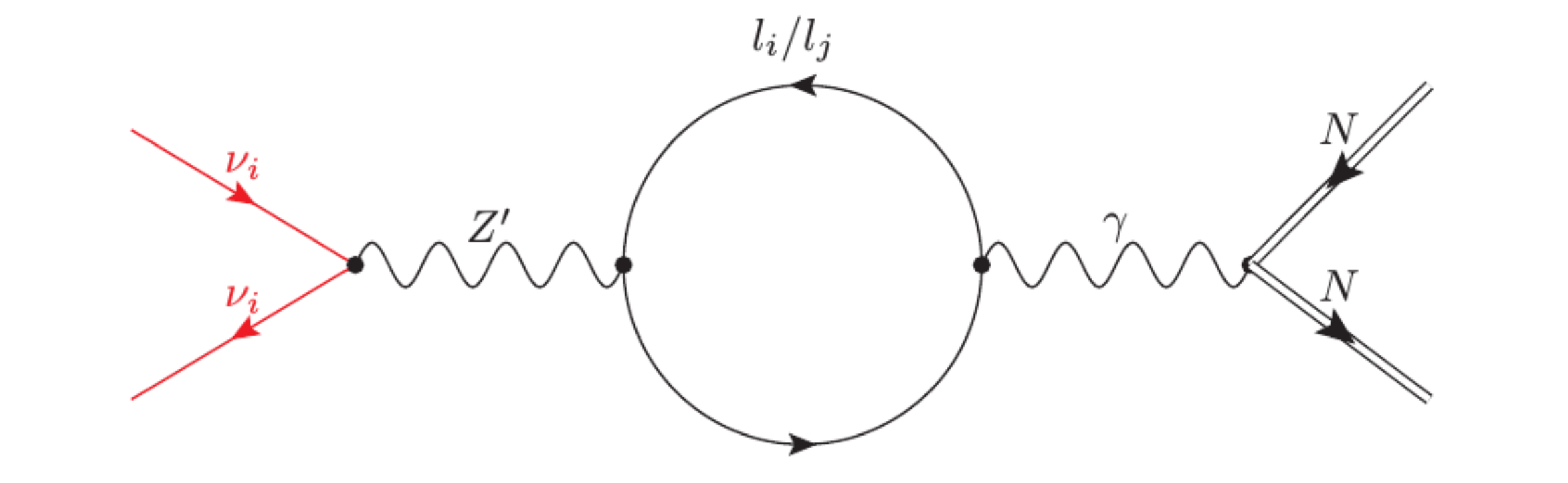}
\caption{}\label{feyn1}
\end{subfigure}%
\caption{(\ref{feyn2})\small \em{ Dominant contribution to spin-independent DM-nuclear scattering and (\ref{feyn1}) the dominant 
channel contributing to neutrino background.}} 
\label{feyn}
\end{figure}

The presence of an extra gauge boson $Z^{\prime}$ in the leptophilic U(1) models can potentially act as a new mediator and open up new annihilation channels in the dark matter scenario, when the dark matter couples to the $Z^{\prime}$. When a DM candidate couples to $Z'$, then resultant DM annihilations to leptons can be interpreted as observed and expected electron (positron) excess in the DM indirect detection experiments. A vector like fermion dark matter~\citep{Bi:2009uj,Arcadi:2018tly,Foldenauer:2018zrz,Altmannshofer:2016jzy} is natural and minimal extension, as it does not contribute anything to gauge anomaly. A vector like fermion $\chi$ (DM Candidate) can be added through a term in the Lagrangian $q_{\chi} g^{\prime} \bar{\chi} Z^{\prime} \chi $, where $q_{\chi}$ is the gauge charge of the VLF under new $U(1)$. Through this term major DM annihilation channels open up namely the s-channel, $\bar{\chi}\, \chi \to Z^{\prime} \to \bar{\nu} \, \nu / \bar{l}\,l$ and the t-channel annihilation through $\bar{\chi} \, \chi \to Z^{\prime} \, Z^{\prime}$ through a t-channel $Z^{\prime}$ propagator. When both the DM and $Z^{\prime}$ are of the similar mass i.e. $m_{DM} \approx m_{Z^{\prime}}/2$, then only the s-channel annihilation dominates. For a light ($ m_{Z^{\prime}} \approx 10 ~$MeV) $Z^{\prime}$, GeV scale DM annihilation mainly happens through the t-channel process. This is evident in the case of extended models containing right handed neutrino dark matter candidates, where possible explanation to $(g-2)_\mu$ anomaly, neutrino trident process and neutrino masses \citep{Patra:2016shz,Baek:2015fea} allows for only very light $Z^{\prime}$ in the $U(1)_{\mu-\tau}$ model. In the context of DM direct detection, due to lack of tree level coupling between quarks and $Z'$ DM-nucleus interactions will be induced at loop level via $Z'$-$\gamma/Z^0$ mixing. Direct-detection of dark matter experiments involve nuclear recoil energies typically less than few hundred keVs. At these order of recoil energies, dominant dark matter nuclear interactions will be mediated via $Z'$-$\gamma$ mixing through the feynman diagram \citep{Altmannshofer:2016jzy,Duan:2017qwj} shown in figure \ref{feyn2}, given by,
\begin{eqnarray}
\delta^{\mu\nu}_{ij}=\frac{1}{(2\pi^2)}[- l^\mu l^\nu + g^{\mu\nu} l^2]\int^1_0 dx\,( \log\frac{x(x-1)l^2 + m^2_{l_i}}{x(x-1)l^2 + m^2_{l_j}})x(1-x)
\end{eqnarray}
where, $l$ is momentum transfer, $m_{l_{i(j)}}$ is mass of i(j)$^{th}$ flavour lepton in  loop \ref{feyn}.

 Similarly neutrino-nuclear interactions mediated by figure \ref{feyn1} will serve as the chief BSM background to dark matter Direct-detection in considered models.
 A scalar DM~\cite{Baek:2019qte} candidate can also be introduced where the gauge anomaly is taken care of by other new particles.

Following different constraints discussed in the Ref.~\citep{Bauer:2018onh}, the limits on $U(1)_{e-\mu}$ ,$U(1)_{e-\tau}$  and $U(1)_{\mu-\tau}$ models are presented here. 
The major constraints on these models come from various beam dump experiments \citep{Riordan:1987aw,Bjorken:1988as,Bross:1989mp}. In the electron beam dump experiments like E137, E141 (SLAC), E774 (Fermilab) etc where electron beam falls on detector material and the dielectric state final state cross section is measured. 
The electron production through light $Z^{\prime}$ decay is possible in the models $U(1)_{e -\tau}, U(1)_{e -\mu} $ where direct $Z^{\prime}$ couplings to the electron are present. 
For models like $U(1)_{\mu - \tau}$ where the light boson couples to the electron only through loop effects, the constraints from the electron beam dump experiments become less stringent. 
For the leptophilic models like these, due to absence of direct quark interaction, cannot be constrained by the proton beam dump experiments. 
Borexino\citep{Bellini:2011rx} and TEXONO\citep{Deniz:2009mu} experiments measure the cross sections of the processes where neutrinos scatter off the electron i.e. the $\nu_{\alpha} - e$ process. 
These processes will be significantly modified where the light $Z'$ couples to the electron along with different neutrinos, while for the $U(1)_{\mu-\tau}$, 
these interaction only happen through a $Z-Z^{\prime}$ mixing, and therefore constraints are less stringent. 
In the neutrino trident \cite{Ballett:2019xoj} production process like $\nu_{\mu} Z \to \nu_{\mu} \mu^{+} \mu^{-} $ which is measured in the neutrino experiments like CCFR, Charm-II \citep{Vilain:1994qy}, nuTEV etc can provide not so suppressed contributions through the light $Z^{\prime}$ for the U(1) models having direct $\mu$ couplings i.e. $U(1)_{\mu - \tau}, U(1)_{\mu - e}$, while the constraint will be way weaker for $U(1)_{e - \tau}$. Presence of new leptonic forces \cite{Wise:2018rnb} can contribute to matter effects for neutrino oscillations. Due to this effect Super-K provides additional constrains for $U(1)_{e-\mu}$ ,$U(1)_{e-\tau}$, while $U(1)_{\mu-\tau}$ remains insensitive.  COHERENT experiment currently only has preliminary CE$\nu$NS measurement which does not put stringent constraints.  In addition to this, for an ultra light $Z'$ (m$_{Z'} \leq 1 $\,eV), constraints derived from astrophysical observations and meson decays  have been studied in Ref. \cite{Dror:2020fbh}.


\section{Neutrino-Nucleus Interaction Rate}
\label{nuNucleusIR}
In context of DM direct detection experiments, incident neutrinos with energies upto tens of MeV can coherently interact with the nucleus of detecting material producing nuclear recoils, which are hard to differentiate against DM nucleus interactions. Due to the weak nature of neutrino interactions, the detectors are impossible to shield against them. Even without the detection of DM candidates, with increased exposure time and incident flux, experiments can detect coherent neutrino nucleus scattering \cite{Akimov:2017ade} and provide us with the opportunity to probe new neutrino physics.

In the process of coherent neutrino-nucleus scattering (CE$\nu$NS) introduced in Ref.~\cite{Freedman:1973yd}, for small momentum transfer i.e. $qR\leq 1$, where $q$ and $R$ are momentum transfer and radius of the target nucleus respectively, the incident neutrino can scatter with the entire nucleus coherently. In general CE$\nu$NS can lead to nuclear recoils upto a few keVs, which in the case of $Xe^{131}$ target can be translated to incident neutrino energies upto $\approx$ 50 MeV. While, in the Standard Model, the interaction is mediated by $Z^0$ boson, with the presence of light $Z^{\prime}$ boson in $U(1)_{i-j}$ model, CE$\nu$NS is further augmented by $Z^{\prime}-Z^0/\gamma$ mixing. In the regime when nucleus recoil energies are at most few hundred keVs, the dominant CE$\nu$NS due to extra $Z'$ boson will be mediated by $Z'-\gamma$ as shown in Fig.~\ref{feyn1}.

\begin{figure}
\centering 
\begin{subfigure}{.49\textwidth}\centering
  \includegraphics[width=\columnwidth]{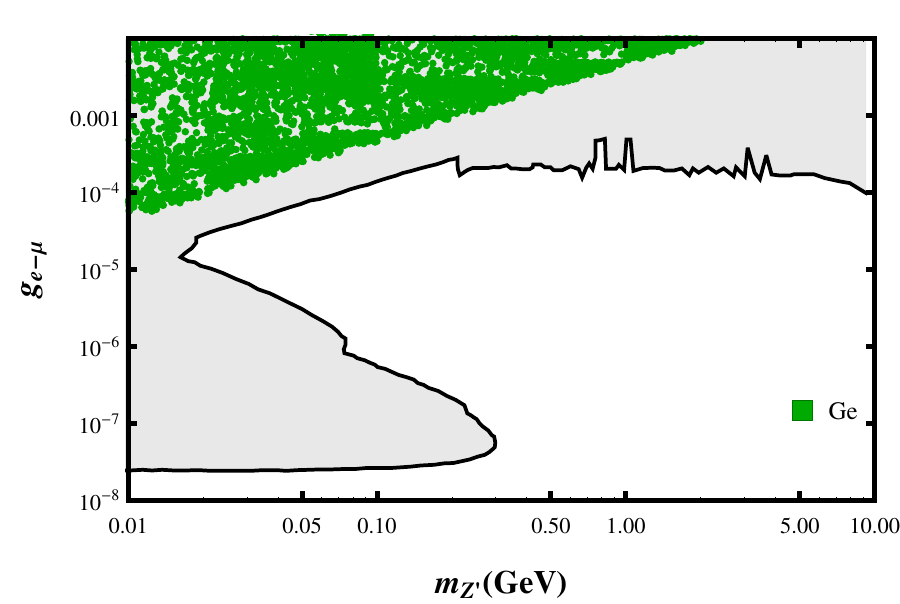}
  \caption{\small \em{Scatter points for $R_{e - \mu}\geq 1.05$ with dis-allowed regions shaded.}}
  \label{ratiog1p05Lemu}
\end{subfigure}%
\begin{subfigure}{.49\textwidth}\centering
  \includegraphics[width=\columnwidth]{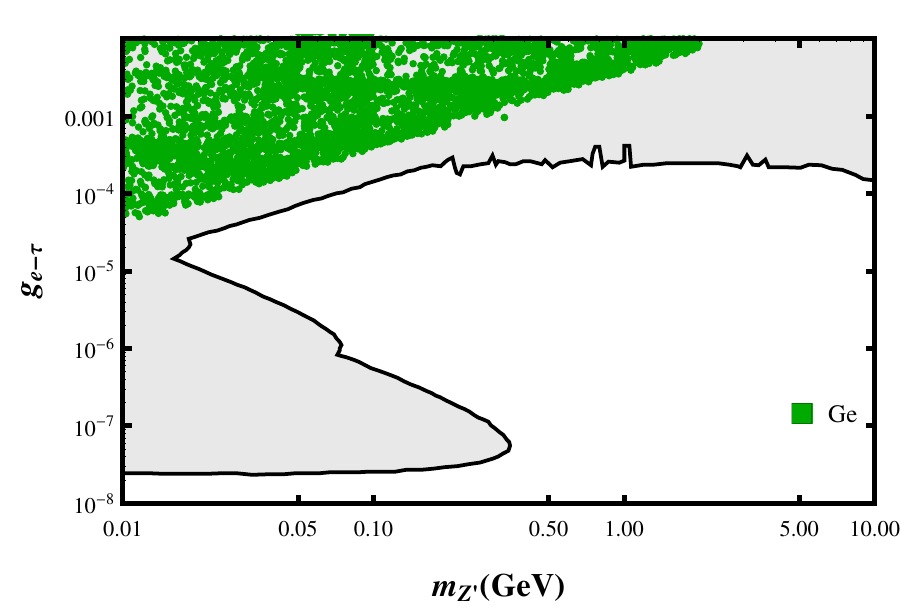}
  \caption{\small \em{Scatter points for $R_{e - \tau}\geq 1.05$ with dis-allowed regions shaded.}}
  \label{ratiog1p05Letau}
\end{subfigure}%
\caption{\small \em{We show parameter regions disallowed (shaded) by experiments \citep{Bauer:2018onh} in  g$_{i-j}$ vs m$_{Z'}$(GeV) planes for U(1)$_{e-\mu}$ ,U(1)$_{e-\tau}$. Green scatter points are measure of CE$\nu$NS enhancement $R_{i-j}$ as specified in each case for Ge$^{68}$. } }
\label{figmodelconstrainsemu}
\end{figure}

\begin{figure}[h!]
\begin{subfigure}{.49\textwidth}\centering
  \includegraphics[width=\columnwidth]{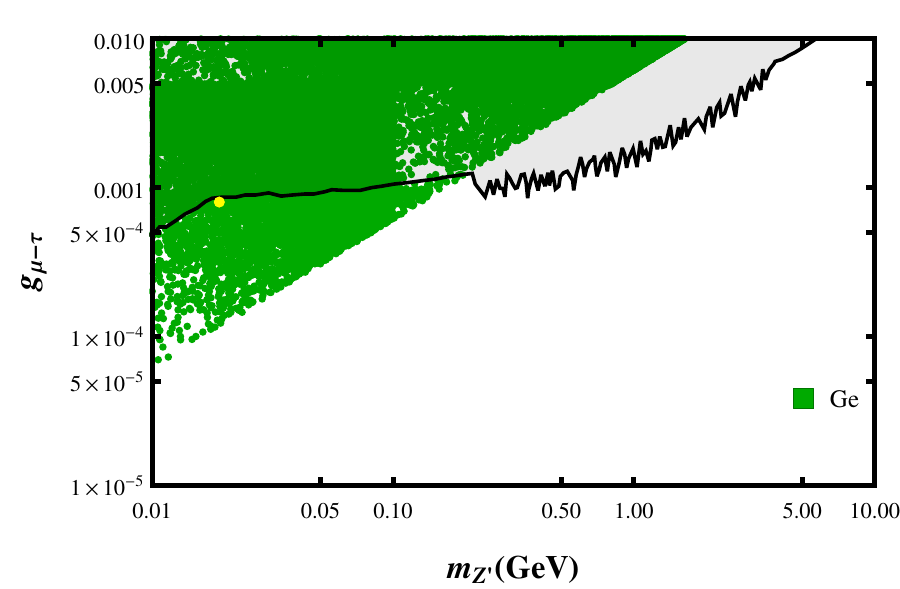}
  \caption{\small \em{Scatter points for $R_{\mu-\tau}\geq 1.05$ with dis-allowed regions shaded.}}
  \label{ratiog1p05Lmutau}
\end{subfigure}%
\begin{subfigure}{.49\textwidth}\centering
  \includegraphics[width=\columnwidth]{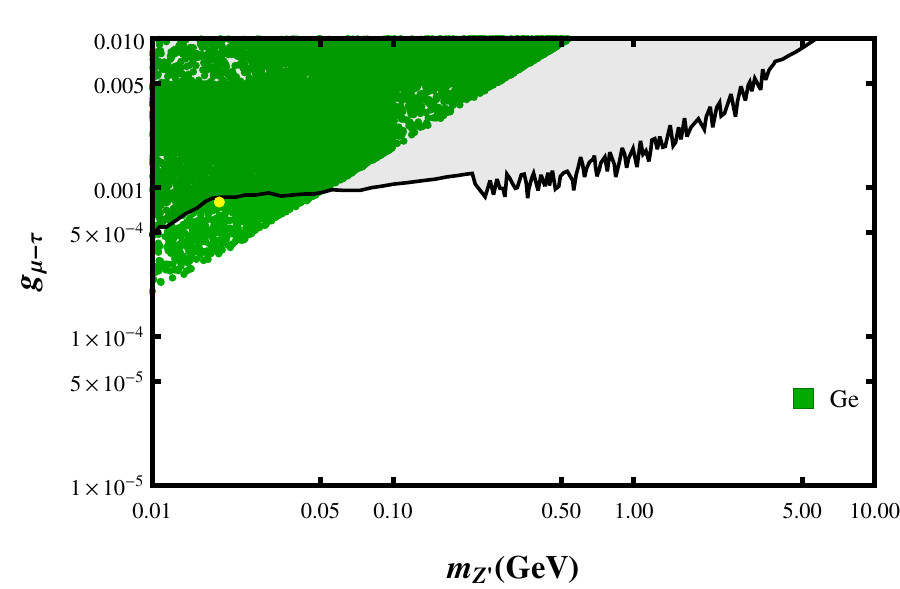}
  \caption{\small \em{Scatter points for $R_{\mu-\tau}\geq1.5$ with dis-allowed region shaded.}}
  \label{ratiog1p5Lmutau}
\end{subfigure}%
\\
\begin{subfigure}{.51\textwidth}\centering
  \includegraphics[width=\columnwidth]{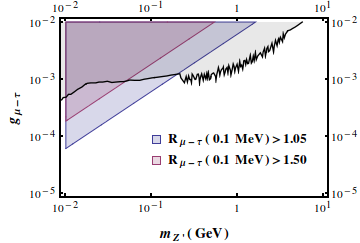}
  \caption{\small \em{Region plot showing different R$_{\mu-\tau}$ with incident neutrino energy 0.1 MeV   }}
  \label{regionp1Lmutau}
\end{subfigure}%
\begin{subfigure}{.51\textwidth}\centering
  \includegraphics[width=\columnwidth]{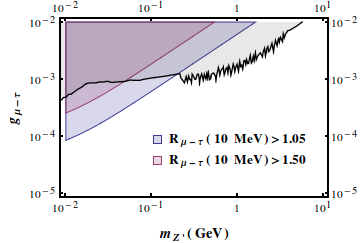}
  \caption{\small \em{Region plot showing different R$_{\mu-\tau}$ with incident neutrino energy 10 MeV  }}
  \label{region10Lmutau}
\end{subfigure}%

\caption{\small \em{We show parameter regions disallowed (shaded greay) by experiments \citep{Bauer:2018onh} in  g$_{i-j}$ vs m$_{Z'}$(GeV) planes for  U(1)$_{\mu-\tau}$ models. In top panels green scatter points are measures of $R_{\mu-\tau}$ as specified in each case for Ge$^{68}$. Bottom panels show region plots for different $R_{\mu-\tau}$ for incident neutrino energy 0.1 MeV (left panel) and 10 MeV (right panel). Benchmark point: $Z'$ mass $m_{Z^{\prime}}$ = 19 MeV and coupling $g_{\mu-\tau} = 8 \times 10^{-4} $, showing maximum enhancement in allowed parameter space of the U(1)$_{\mu-\tau}$ model is highlighted in top panels.} }
\label{figmodelconstrains}
\end{figure}

Taking into account the effects of the mixing, the total neutrino-nucleus differential scattering cross-section in $U(1)_{i - j}$ can be written as
 \begin{eqnarray}
 \label{differentialscattering}
&&\frac{d\sigma_{i-j}}{dE_r}=\frac{d\sigma_{SM}}{dE_r}
-\frac{{m_N}\,G_f\, Q_{\nu N{i-j}} Q_{\nu N} \left(1-\frac{{E_r} {m_N}}{2 {E_\nu}^2}\right)F^2(E_r)}{\sqrt{2} \pi  \left(2 {E_r} {m_N}+{m_{Z^{\prime}}}^2\right)}+\frac{{m_N}\, Q_{\nu N{i-j}}^2 \left(1-\frac{{E_r} {m_N}}{2 {E_\nu}^2}\right)F^2(E_r)}{2 \pi  \left(2 {E_r} {m_N}+{m_{Z'}}^2\right)^2},
\end{eqnarray}
where as the SM counterpart for the neutrino-nucleus scattering process is given by, 
\begin{eqnarray}
&&\frac{d\sigma_{SM}}{dE_r}=G_f^2 \frac{m_N}{4 \pi } Q^2_{\nu N} \left(1-\frac{E_r m_N}{2 {E_\nu}^2}\right)F^2(E_r).
\end{eqnarray} 
Here $G_f$ is the Fermi constant, $Q_{\nu N} = N - (1-4\sin^2\theta_w) Z$ is effective weak hyper-charge in the SM for the target nucleus with $N$ neutrons and $Z$ protons and $F(E_r)$ is the Helm form factor given in Ref.~\cite{Lewin:1995rx}, that exhibits the loss of coherence above recoil energies of $\approx$ 10 keV. The effective weak interaction vertex in the neutrino part for the BSM case of $U(1)_{i -j}$ model can be written as,
\begin{eqnarray}
 Q_{\nu N{i-j}} =  g_{i-j}^2\frac{2\, \alpha_{EM}}{\pi} \delta_{ij} Z 
\end{eqnarray}
where $ g_{i-j}$ is coupling given in Eq.~\ref{intLij},  $\alpha_{EM}$ is the fine structure constant and $\delta_{ij}$ is the scalar part of loop factor given in equation \ref{intLij}.

\begin{figure}[h!]
\centering 
\begin{subfigure}{.49\textwidth}\centering
  \includegraphics[width=\columnwidth]{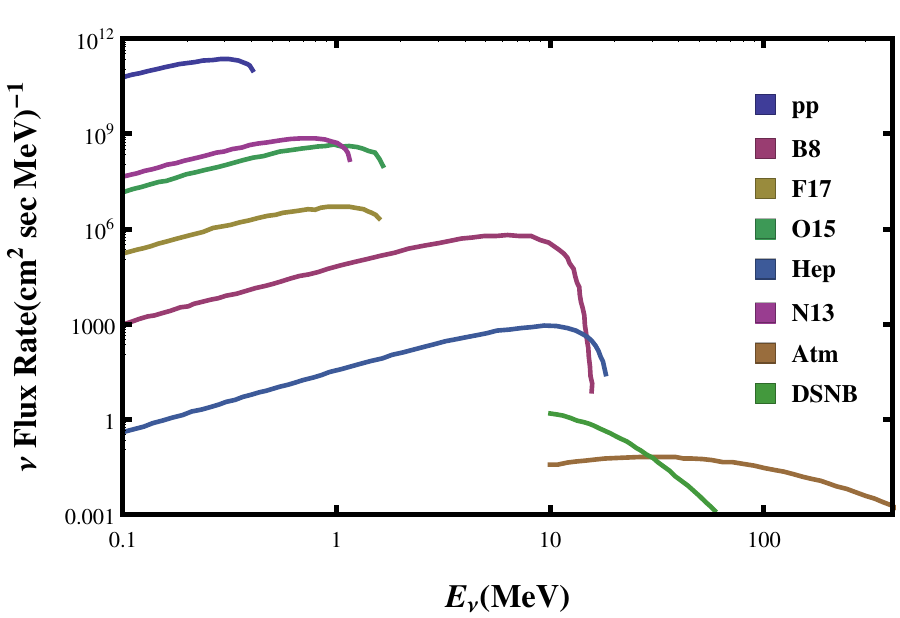}
  \end{subfigure}%
\caption{\small \em{ Relevant continuous neutrino sources. Solar: pp, b8, F17, O15, Hep, N13; Atmospheric: Atm; Diffuse supernova neutrino background: DSNB}}
\label{neutrinoflux}
\end{figure}

 In order to study possible modification of neutrino background in the U(1)$_{i-j}$ models, we check the significance of BSM effects by looking for a variation in the CE$\nu$NS rate compared to the SM one, against the parameter region allowed by the experiments discussed in Ref.~\citep{Bauer:2018onh}. To discern the beyond standard model effect of these models, we define a ratio,
\begin{eqnarray}
R_{i-j}=\frac{\sigma_{i-j}}{\sigma_{SM}}= \frac{\int^{E_r^{max}}_0 \frac{d\sigma_{i-j}}{dE_r}d E_r}{\int^{E_r^{max}}_0 \frac{d\sigma_{SM}}{dE_r} d E_r} , 
\label{ratioxs}
\end{eqnarray} 
where $E_r ^{max}\approx \frac{2(E_\nu)^2}{m_N}$. The allowed parameter space from different constraints along with the increase of CE$\nu$NS rate for these models are presented in Fig.~\ref{figmodelconstrainsemu} for the case of U(1)$_{e-\mu}$ and U(1)$_{e-\tau}$ models  and Fig.~\ref{figmodelconstrains} for the case of U(1)$_{\mu-\tau}$ model.
For computing $R_{e-\mu}$ and $R_{e-\tau}$ in Fig.~\ref{figmodelconstrainsemu}  , incident $\nu_e$ is considered while $\nu_\mu$ is considered for $R_{\mu-\tau}$ calculation, presented in Fig.~\ref{figmodelconstrains} . The $\nu$ flavors are chosen such as to maximize the enhancement in $R_{i-j}$ as considering $\nu_j$ for calculating $\sigma_{i-j}$ leads to diminution of $R_{i-j}$ values. Incident neutrino energy range is chosen from 0.1 MeV to 100 MeV. The considered energy range and flavor composition are influenced by the relevant neutrino background sources (see Fig.~\ref{neutrinoflux}) .

The grey shaded regions signify the parameter space in  $g_{i-j}$ vs $m_{Z^{\prime}}$ plane ruled out by the experiments. 
The green dots are combination of $( g_{i-j}$, $m_{Z'})$ corresponding to the enhancement in the CE$\nu$NS rate as described by the quantity $R_{i-j}$ in each case. 
In all the panels, it is observed that with decreasing mass of $Z'$, it becomes possible to attain same  $R_{i-j}$ ratios at lower values of $ g_{i-j}$. 
This can be attributed to the effect from the interference term in total $\sigma_{i-j}$ which is proportional to $\frac{g^2_{i-j}}{(2E_r m_N + m^{2}_{Z'})}$. 
Fig.~\ref{ratiog1p05Lemu} and Fig.~\ref{ratiog1p05Letau} in the top panel represent the parameter space ruled out due to combined constraints from different experiments, respectively for the $U(1)_{e-\mu}$ and $U(1)_{e-\tau}$ models. 
The green dots are combinations of $( g_{i-j}$, $m_{Z'})$ such that $ R_{i-j} \geq 1.05$ in each case.  For the case of $ U(1)_{e-\mu}$ and $ U(1)_{e-\tau}$ models, it is observed that the points with 5\% or more enhancement in the CE$\nu$NS lie in the shaded region, leading us to decipher that even 5\% increment is not possible due to the BSM effects within the allowed  parameter space. On the other hand, for the $U(1)_{\mu-\tau}$ model in Fig.~\ref{ratiog1p05Lmutau} and Fig.~\ref{ratiog1p5Lmutau}, green dots signify the combination of $( g_{i-j}, m_{Z'})$ such that $R_{\mu-\tau} \geq 1.05$ and $R_{\mu-\tau} \geq 1.5$ respectively.  Subsequently  we show region plots for $R_{\mu-\tau}\geq$ 1.05 and $R_{\mu-\tau}\geq$ 1.50  for incident neutrino energies 0.1 MeV and 10 MeV respectively in Fig.~\ref{regionp1Lmutau} and Fig.~\ref{region10Lmutau}  .

For the $U(1)_{\mu-\tau}$ model, CE$\nu$NS enhancement of as high as 50\% can be achieved in the allowed region.
This specific nature can be attributed to the relaxation of constraints from experiments involving $\nu_e$ and electron, as $Z'$ does not have tree level couplings with $\nu_e, e^{\pm}$ in $U(1)_{\mu-\tau}$ model. 
Furthermore, the COHERENT experiment which has measured CE$\nu$NS, has threshold recoil energy in the vicinity of 5 keV for CsI, Ge and Xe targets \cite{Akimov:2015nza}. 
This translates to neutrino incident around 10 MeVs, therefore unable to constrain parameter space for lower $\nu$ energies. Furthermore, beyond standard contribution to CE$\nu$NS  is proportional to $\frac{g^2_{i-j}}{(2E_r m_N + m^{2}_{Z'})}$. 
This contribution  scales as $\frac{g^2_{i-j}}{m^2_{Z'}}$ at heavier $Z'$ but below  $m_{Z'}\leq \sqrt{2 E^{th}_r m_{N_{Ge}}}\approx$ 35 MeV, the contribution only scales as  $g^2_{i-j}$. Therefore current COHERENT constraints on relevant parameter space remains weak. However next phase of COHERENT experiment with significant lower threshold around 10 eV \citep{Billard:2018jnl} can scan the relevant parameter space more strongly.
Due to the enhancement seen in parameter space allowed by experiments only for the case of $U(1)_{\mu-\tau}$ model  , we plan to dig deeper only into the $U(1)_{\mu-\tau}$ model in the following discussion.

 As discussed previously, CE$\nu$NS can lead to measurable nuclear recoils in detectors. 
 A regular neutrino flux would lead to detection of scattering events over a time depending on luminosity of incident neutrinos and strength of the interaction. 
 The neutrino-nucleus event rate equation which determines the neutrino matter interaction, 
 can be written as~\cite{Gonzalez-Garcia:2018dep}

\begin{eqnarray}
\label{ratenu}
\frac{d R_{\nu-N}}{d E_r} = \frac{\epsilon}{m_N} \int_{E^{min}_\nu} \mathcal{A}(E_r) \left.\frac{d \phi_\nu}{d E_{\nu}}\right|_{\nu_\alpha} P(\nu_\alpha \rightarrow\nu_\beta,E_\nu) \frac{d \sigma (E_\nu,E_r,\nu_\beta)}{d E_r} d E_\nu
\end{eqnarray}
here $\epsilon$ is the exposure of the experiment measured in units of mass $\times$ time, $\mathcal{A}(E_r)$ is the detector efficiency and is set to one in following calculations. $E^{min}_\nu$ is minimum incident neutrino energy required to produce a detectable recoil for a material nucleus of mass $m_N$ with energy $E_r$, which in the limit of $m_N >> E_\nu$ can be written as,
\begin{eqnarray}
E^{\rm min}_\nu = \sqrt{\frac{m_N E_r}{2}}.
\end{eqnarray}

\begin{figure}[h!]
\centering 
\begin{subfigure}{.49\textwidth}\centering
  \includegraphics[width=\columnwidth]{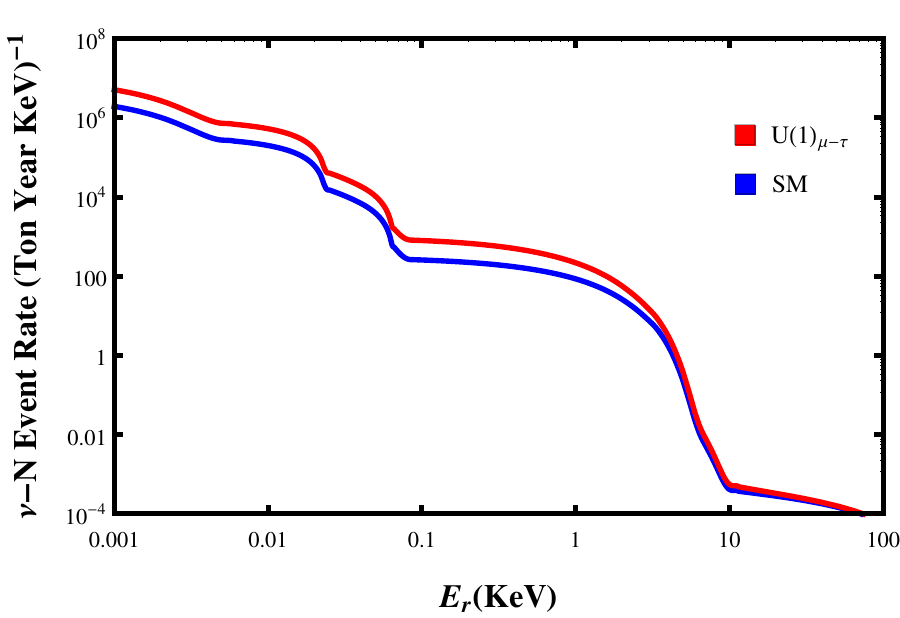}
  \caption{ }
  \label{rateGe}
\end{subfigure}%
\begin{subfigure}{.49\textwidth}\centering
  \includegraphics[width=\columnwidth]{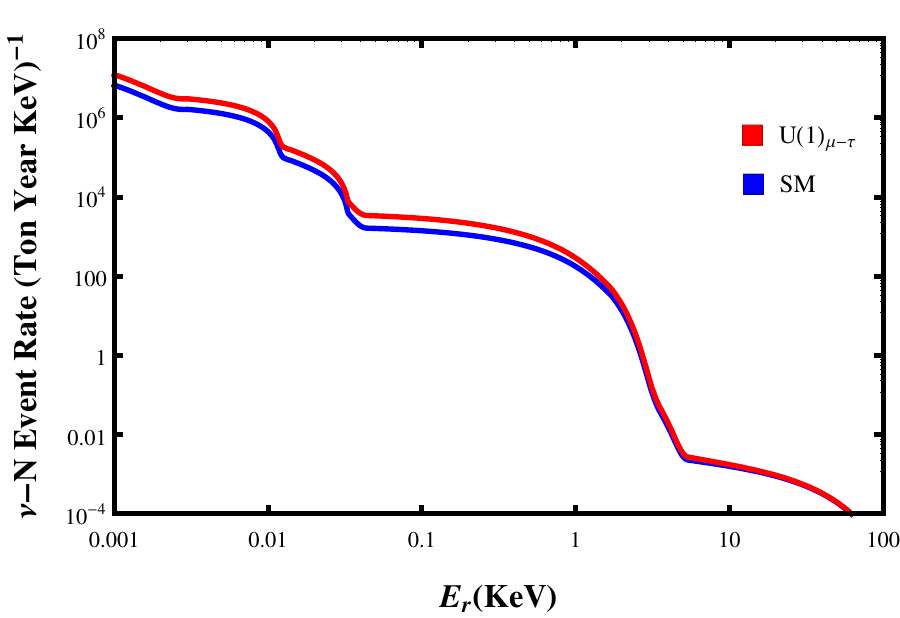}
  \caption{}
  \label{rateXe}
\end{subfigure}%
\caption{Event rate for neutrino-nucleus scattering with change of recoil energy. Blue line is the event rate for SM whereas red line is for $U(1)_{\mu-\tau}$. Left panel is  for Ge$^{68}$ and right for Xe$^{131}$. Benchmark model parameter space for these plots: $Z'$ mass 19 MeV and coupling $g_{\mu-\tau} = 8 \times 10^{-4} $.}
\label{figeventrate}
\end{figure}

Here, $\frac{d \sigma (E_\nu, E_r, \nu_\beta)}{d E_r}$ is $\beta$ flavor dependent neutrino-nucleus differential scattering cross-section and $\left.\frac{d \phi_\nu}{d E_{\nu}}\right|_{\nu_\alpha}$ is  the incoming  neutrino flux of flavor $\alpha$. The fluxes used in this analysis involve fluxes from solar, atmospheric, diffuse supernova neutrinos, which can be found in  Refs.~\cite{Strigari:2009bq,Billard:2013qya} and have been redrawn in Fig.~\ref{neutrinoflux}. 
Apart from the continuous sources, electron capture on Be$^7$ leads to two mono-energetic neutrino lines at 384.3 keV and 861.3 keV and have been taken into account. $P(\nu_\alpha \rightarrow\nu_\beta,E_\nu)$ is the transition probability of $\nu_\alpha \rightarrow \nu_\beta$ in the incident flux. Electron neutrinos emitted from different layers of solar core can undergo flavor oscillations in the inter-lying medium, therefore leading to finite probability of incident solar neutrinos to be of different flavor when they reach earth. 
It was shown in Refs.~\citep{Hernandez:2010mi,Lopes:2013nfa} that survival probability of neutrinos with particular flavor remain very close to each other for two or three flavor neutrino oscillation. Therefore, we use the neutrino survival probabilities for two flavor neutrino oscillation model studied in Ref.~\cite{Lopes:2013nfa} to calculate $P(\nu_e \rightarrow\nu_\mu,E_\nu)$ which is used to compute neutrino-nucleus rate equations and afterwards, the neutrino floor, for $U(1)_{\mu-\tau}$.

  Using the rate Eq.~\ref{ratenu}, we show in Fig.~\ref{figeventrate}, dependence of neutrino-nucleus scattering rate on recoil energy, with an exposure of 1 ton year. The contours represent the number of CE$\nu$NS events per keV of nuclear recoil energy in one ton detector of given material, counted over a year. The incident neutrino flux rate is the most drastically changing function in the integrand, leading the profile of contours to mimic it. As can be seen in Fig.~\ref{neutrinoflux}, the total neutrino flux rate experiences a big drop with increase in incident neutrino energy. In comparison to the solar neutrino flux, very little is contributed by the atmospheric and DSNB neutrino sources which contribute beyond $E_{\nu} \sim $20 MeV. 
Similar profile is seen in event rate contours in Fig.~\ref{figeventrate}. The bulges appearing in the event rate contours can be attributed to switching off of individual neutrino flux sources in the total flux. As an example, the first two bulges seen at 0.003 and 0.023 keV recoil energies in the case of Germanium nuclei can be sourced to PP spectrum and Be$^7$ 861 keV line. In the left panel~\ref{rateGe}, the event rate for SM is shown along with the event rate for $U(1)_{\mu-\tau}$ model for Ge based detectors. Similarly, event rates for both the models are shown in the right panel~\ref{rateXe} for Xenon based detectors. Enhancement by factor around 2.8 can be seen in the case of Germanium and by a factor of 1.8 for Xenon for recoil energies of sub-keV regime. Beyond 1 keV the enhancement diminishes rapidly as momentum transfer increases beyond the chosen $m_{Z'}$. For further details regarding enhancement in CE$\nu$NS event rates see table
\ref{table:1} and \ref{table:2}.

\section{Neutrino Floor}
\label{sectionneutrinoFloor}

In the context of DM direct detection experiment, {\it neutrino floor} represents the neutrino background to the DM signal events. The projection of background CE$\nu$NS events in terms of signal DM parameter space is enshrined through the neutrino floor. 
Neutrino floor is defined as the minimum value of DM-nucleon scattering cross-section, below which nuclear recoil due to DM will remain indistinguishable from the those recoils due to neutrinos. The cross section on the neutrino floor will be set  such that for each DM mass, the ratio of 2.3 DM signal events (90\% C.L.)\citep{Billard:2013qya} to one neutrino background event is maintained. 
This can lead us to establish a boundary in DM-nucleon scattering cross-section above which there is certainty ( at 90\% C.L.) that the observed events, if any, are indeed the DM signal events, i.e. they are coming from DM-nucleon interactions. 

Following the Refs.~\cite{Read:2014qva,Pato:2015dua} investigating local DM, it is becoming increasingly certain that DM also permeates our immediate galactic vicinity. 
Recent constraints~\cite{deSalas:2019rdi} estimate the local DM density $\rho_{DM}\simeq 0.3-0.4 \rm GeV/cm^3$. 
When the DM particle passes through the matter, it can interact with constituents of the atom. 
These interactions can lead to elastic or inelastic scattering with electrons and elastic scattering with nucleus, depending on the scale of momentum transfer and nature of DM interactions with matter. 
If the DM matter interactions take place inside a detector then they can be detected by measuring recoiling energy of nucleus or electron. 
For DM of mass greater than few hundred MeVs, DM-nucleus scattering plays a more important role in detection of DM \cite{Wyenberg:2018eyv}. 
The differential DM-nucleus scattering event rate is given by
\citep{Undagoitia:2015gya}.
\begin{equation}
\label{rateDM}
\frac{d R_{ DM-N}}{d E_r} = \epsilon\,\frac{\rho_{DM}\sigma^0_n A^2}{2m_{DM}\mu^2_{n}}F^2(E_r) \int_{v_{min}} \frac{f(v)}{v } d^3 v
\end{equation}
Here $\epsilon$ is the exposure of the detector given in units of MT (mass$\times$time), $m_{DM}$ is the DM mass $\mu_{n}$ is DM-nucleon reduced mass, $A$ is the mass number of target nuclei, $\sigma^0_n$ is the DM-nucleon scattering cross-section at zero momentum transfer. $F(E_r)$ is the Helmholtz form factor. 
The Maxwell-Boltzmann distribution function, $f(v)$ is assumed to describe the velocity distribution of DM in Earth frame and $v_{min}=\sqrt{m_N E_r/2\mu^2_N}$ where $\mu_N$ is DM-nucleus reduced mass. The Integral in Eq.~\ref{rateDM} can be calculated analytically as \citep{Lewin:1995rx}
\begin{eqnarray}
\int_{v_{min}} \frac{f(v)}{v}d^3 v = \frac{1}{2v_0 \eta_{E}}\left[erf(\eta_+)-erf(\eta_-)\right]- \frac{1}{\pi v_0 \eta_E }\left(\eta_+-\eta_-\right)e^{\eta^2_{esc}}
\end{eqnarray}
Here $\eta_E = \frac{v_E}{v_0}$,$\eta_{esc} = \frac{v_{esc}}{v_0}$ and $\eta_\pm = min\left(\frac{v_{min}}{v_0}\pm \eta_E,\frac{v_{esc}}{v_0}\right)$, where $v_0$ is local galactic rotational velocity, $v_{E}$ velocity of Earth with respect to galactic center, $v_{esc}$ escape velocity of DM from galaxy. We have used values  $v_0=220$km/s, $v_E=$232 km/s and $v_{esc}=$ 544 km/s  in above calculations. 

To construct the neutrino floor, first the exposure required to produce one neutrino event needs to be evaluated. 
That is done following Eq.~\ref{ratenu} and then setting $\int^{E_r^{max}}_{E_{th}}\frac{dR}{dE_r}dE_r=1$. 
In this integral, the minimum recoil energy is taken as the threshold energy $E_{th}$ and maximum nuclear recoil energy, $E_r^{max}$ is chosen to be 100 keV. 
To put it in an alternate way, the mass of the detector (M) times the time for which the experiment is run (T) is computed for a given threshold energy such that it gives us exactly $n_\nu$ counts for neutrino scattering events. 
The exposure is expressed as,
\begin{eqnarray}
\epsilon_{n_\nu} = \frac{n_\nu}{1}\left(\int^{E_{max}}_{E_{th}}\frac{1}{m_N} \int_{E^{min}_\nu} \left.\frac{d \phi_\nu}{d E_{\nu}}\right|_{\nu_\alpha} P(\nu_\alpha \rightarrow\nu_\beta,E_\nu) \frac{d \sigma (E_\nu,E_r,\nu_\beta)}{d E_r} d E_\nu\right)^{-1}, 
\end{eqnarray}
where $n_\nu =1$ can be set for one neutrino-nucleus scattering event. 
 
 Next, we use the computed exposure in the dark matter side. The DM-nucleus event rate in Eq.~\ref{rateDM} is integrated through $\int^{E_{DM}^{max}}_{E_{th}}\frac{dR_{DM-N}}{dE_r}dE_r=2.3$ to produce 2.3 DM scattering events, with the same exposure which was required for single neutrino scattering event. That equation can be solved for DM-nucleon scattering cross-section $\sigma^0_n$, using the same threshold for recoil energy lower limit. 
\begin{figure}[h!]
\centering 
\begin{subfigure}{.49\textwidth}\centering
  \includegraphics[width=\columnwidth]{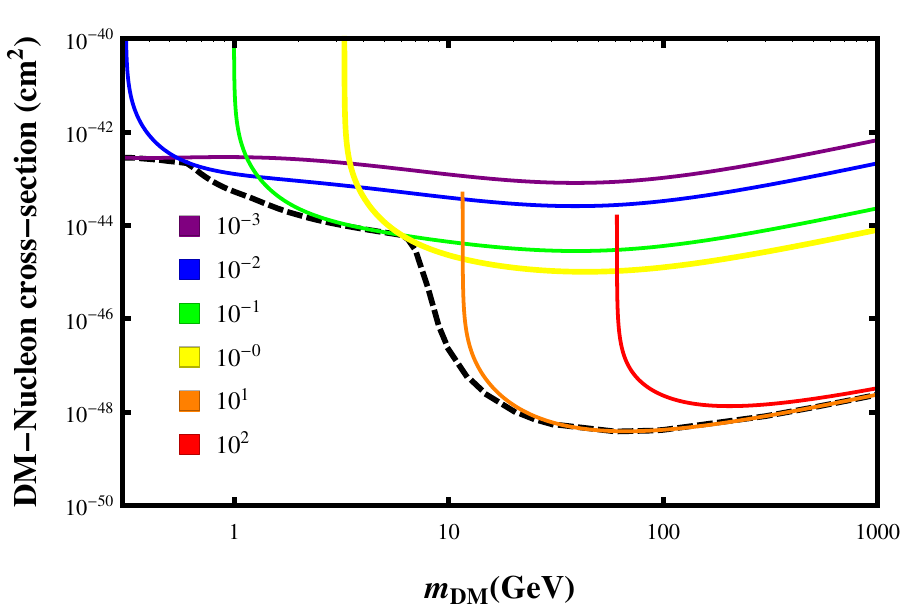}
  \end{subfigure}%
\caption{\small \em{ Black dashed line signify neutrino floor in case of SM with Germanium detector, which is constructed by taking lower limit of $\sigma^0_n$ with varying threshold in logarithmic steps from 0.001 to 100 keV with exposure to attain one neutrino scattering event each. As an example we also show colored $\sigma^0_n$ contours for threshold energies $10^{-3}$, $10^{-2}$, $10^{-1}$, $10^{0}$, $10^{1}$, $10^{2}$ keV highlighting how neutrino floor is spanned.}}
\label{neutrinospan}
\end{figure}
This can be recapitulated in form of the master equation,
\begin{eqnarray}
\label{Nflms}
\int^{E_{DM}^{max}}_{E_{th}}\frac{dR_{DM-N}}{dE_r}dE_r =   \frac{2.3}{1} \int^{E_r^{max}}_{E_{th}}\frac{dR}{dE_r}dE_r \nonumber ,
\end{eqnarray}
that translates to the required DM-nucleon scattering cross-section,
\begin{eqnarray}
\label{Nflcalc}
\sigma^0_n =&&  \frac{2.3}{1}\left(\int^{E_{max}}_{E_{th}}\frac{1}{m_N}\int_{E^{min}_\nu} \left.\frac{d \phi_\nu}{d E_{\nu}}\right|_{\nu_\alpha} P(\nu_\alpha \rightarrow\nu_\beta,E_\nu) \frac{d \sigma (E_\nu,E_r,\nu_\beta)}{d E_r} d E_\nu\right)\nonumber
\\
&&\times\left(\frac{\rho_{DM} A^2}{2m_{DM}\mu^2_{n}}\int^{E^{max}_{DM}}_{E_{th}}F^2(E_r) \int_{v_{min}} \frac{f(v)}{v } d^3 v\right)^{-1}
\end{eqnarray}
Here $E^{max}_{DM}$ is the maximum recoil energy of DM with mass $m_{DM}$ can produce in a given nucleus. It is written as, 

$ 2\, m_{DM} \left(\frac{m_N m_{DM} } {(m_N + m_{DM})}\right) v_{esc}^2$.

 \begin{figure}[h!]
\centering
\begin{subfigure}{.49\textwidth}\centering
  \includegraphics[width=\columnwidth]{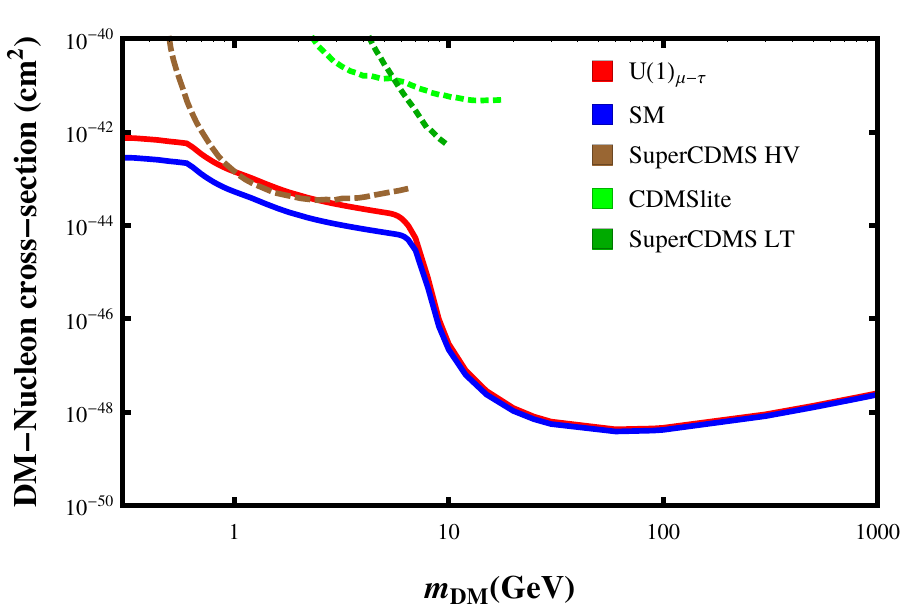}
  \label{NfloorGe}
\end{subfigure}%
\begin{subfigure}{.49\textwidth}\centering
  \includegraphics[width=\columnwidth]{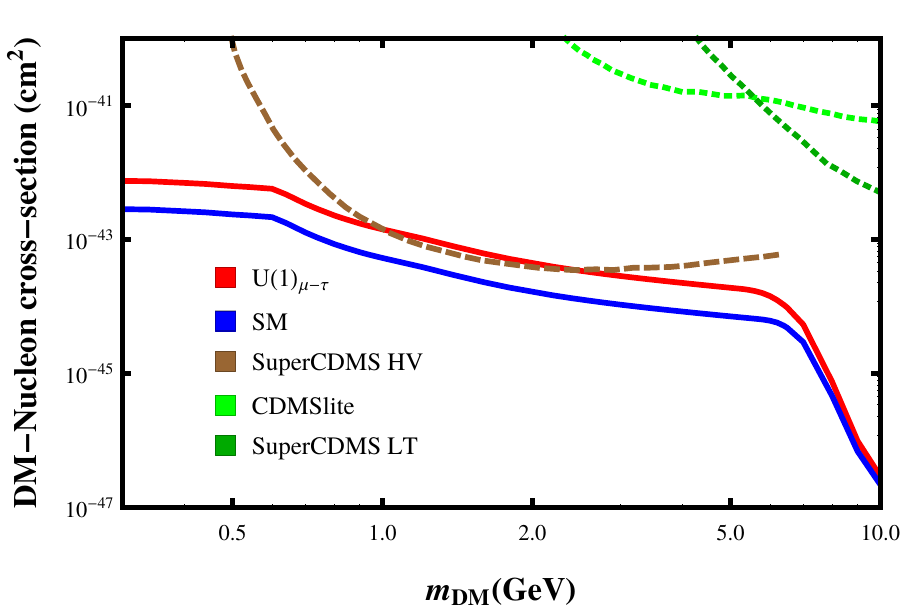}
  \label{NfloorXe}
\end{subfigure}%
\\
\begin{subfigure}{.49\textwidth}\centering
  \includegraphics[width=\columnwidth]{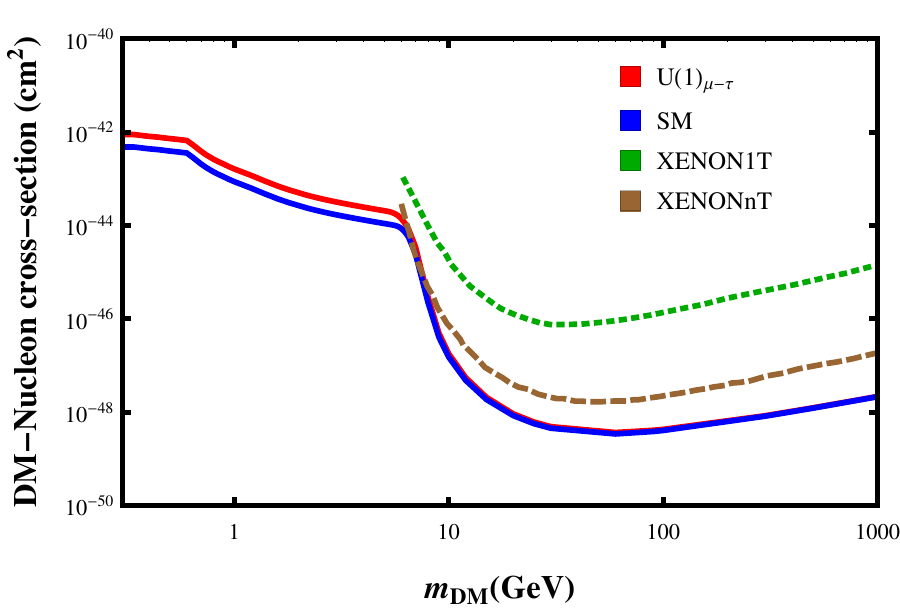}
  \label{NfloorGe10}
\end{subfigure}%
\begin{subfigure}{.49\textwidth}\centering
  \includegraphics[width=\columnwidth]{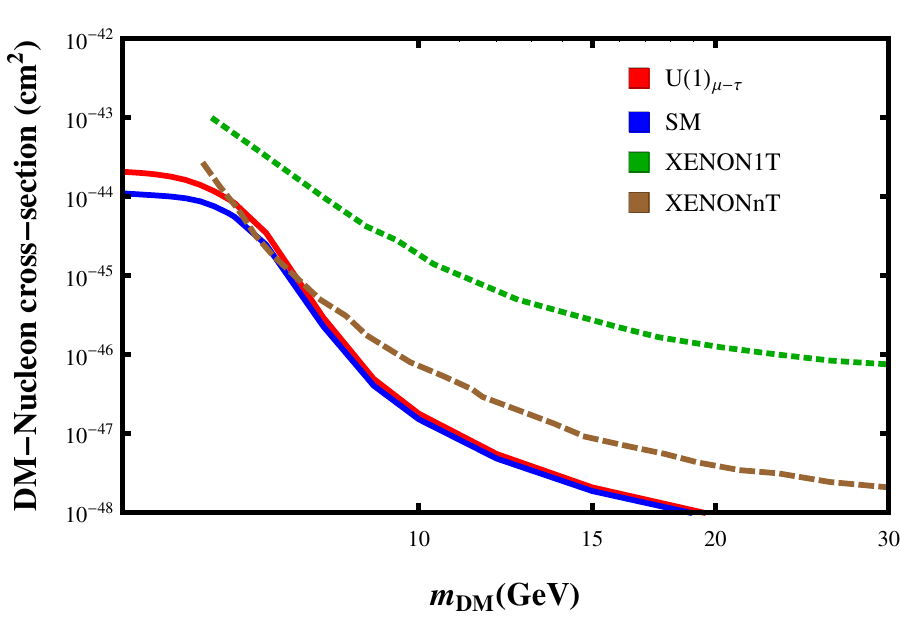}
  \label{NfloorXe}
\end{subfigure}%
\caption{\small \em{Neutrino floor projected in the $\sigma^0_n$ vs $m_{DM}$ plane. Comparison of the neutrino floor for the SM (presented by the blue line) and that for $U(1)_{\mu-\tau}$ (presented by the red line). For different detector materials, Ge$^{68}$ (top panel) and Xe$^{131}$ (bottom panel), dashed and dotted lines respectively show current and future DM-nucleon direct detection exclusion plots. Benchmark chosen for these plots: $Z^{\prime}$ mass 19~MeV and coupling $g_{\mu-\tau} = 0.0008$.}}
\label{Nflr}
\end{figure}

 Using the expression in Eq.~\ref{Nflcalc}, a number of curves for $\sigma^0_n$ (DM-nucleon scattering cross-section) as a function of DM mass are generated with varying threshold energy in logarithmic steps from 0.001~keV to 100~keV. The exposure is kept so that it can generate one neutrino scattering event in each case i.e $n_{\nu} = 1$. Then the lowest cross-section among different $E_{th}$ plots are taken for each DM mass to draw a line in the DM-nucleon cross-section $\sigma^0_n$ versus $m_{DM}$ plane. This curve will put a lower limit on DM-Nucleon cross-section above which we can be certain (at 90 \% C.L.) that the measured events will occur due to DM-nucleon scattering i.e. they are DM signal events. In Fig.\ref{neutrinospan}, we have shown how different threshold energy plots are used to obtain the neutrino floor. When the recoil energy in the DM-nucleon scattering events are smaller than the threshold energy, $E_{th}$, they do not register as recoil events in the detector. If it is assumed that all DM follow same velocity distribution, a lighter DM produces lower recoils. Therefore, with a higher threshold energy, lighter DM recoils remain unnoticed, leading to less sensitivity of the  $\sigma^0_n$ curves to lighter DM. 
 
 With the methodology discussed above, we show in Fig.~\ref{Nflr}, theoretically estimated neutrino floor curves along with current and future sensitivity of different DM direct detection experiments on the $\sigma_n^0 - m_{DM}$ plane. 
Solid lines signify the contours such that for each DM mass, above that cross section, DM scattering events can be differentiated from the neutrino scattering events at 90\% confidence level (i.e. 2.3 DM events per one neutrino events). 
 Plots in the top show, for Germanium based DM direct detection experiments, two neutrino floor being drawn for the SM and $U(1)_{\mu-\tau}$ models, where significant enhancement of the neutrino floor is observed for the BSM case. In top right panel, we zoom in to show $\sigma^0_n$ versus $m_{DM}$ contours for Germanium, with $m_{DM}$ being limited to a range 0.2 to 10 GeV, focusing on the enhancement in $U(1)_{\mu-\tau}$. Almost a consistent enhancement by a factor of 2.7 in the neutrino floor is observed for mass range less than 7 GeV. In this DM mass region, the limit on $\sigma^0_n$ is sensitive to the threshold energies below 1~keV. With that $E_{th}$, lower limit of recoil energies hover around 1~keV or less. As shown in Fig.~\ref{figeventrate}, lower recoil energy contributions are higher and therefore dominant in the $E_r$ integral which lead to a lower $E_{th}$ being translated to lower $E_r$ in our case. For $E_r$ values less than 1~keV, the neutrino-nucleus interaction rate gets enhanced by a factor of 2.7 which eventually translates to an increase of neutrino floor in the sub-10~GeV $m_{\rm DM}$ region by the same factor. 
  Exclusion plots for Germanium based experiments include direct detection reach from projected SuperCDMS HV~\cite{Agnese:2016cpb} experiment, that from CDMSlite~\cite{Agnese:2017jvy} (SuperCDMS LT) experiments shown in top row plots of Fig.~\ref{Nflr} through dashed lines of different colors. 

Bottom panels show graphs for Xenon based experiments, where moderate enhancement of the neutrino floor is observed for the BSM case, by a factor of 1.82 in the neutrino floor for lower mass region. Different dashed lines show DM-nucleon direct detection exclusion plots from projected XENONnT~\cite{Aprile:2015uzo} experiment, and that from the XENON1T experiment, presented in different colors. 
In bottom right panel, we again show $\sigma^0_n$ versus $m_{DM}$ contours for Xenon, with $m_{DM}$ varying in the range 5 to 30 GeV, highlighting the enhancement.  
DM mass going beyond 7 GeV, the neutrino floor starts to show diminishing enhancement. This can be attributed to decreasing augmentation in $U(1)_{\mu-\tau}$ CE$\nu$NS event rate with respect to the SM at higher recoil energies. As discussed earlier, neutrino floor is spanned by taking the lower limit on DM exclusion plots drawn using Eq.~\ref{Nflcalc} by varying threshold energy. The DM mass range of 7 - 15 GeV in the neutrino floor is spanned by varying threshold recoil energies from 1 to 10 keV, which shows not so significant enhancement in CE$\nu$NS rate, as can be seen in Fig.~\ref{figeventrate}. See table \ref{table:3} and \ref{table:4} for further details on neutrino floor enhancement.

It is worthwhile to note that future projection of the exclusion plots from SuperCDMS HV~\cite{Agnese:2016cpb} and XENONnT~\cite{Aprile:2015uzo} experiments have an overlap with the modified neutrino floor in the $U(1)_{\mu- \tau}$ model. The enhancement in the neutrino floor will enable to observe neutrino signal events in these detectors, even in the absence of any DM signal. These events due to the overlap could have been erroneously attributed to DM-nucleon scattering, which are CE$\nu$NS events in reality. Any future signal in that range should be probed with more vigor and from alternative experiments to ascertain the presence of DM. If DM is not present, then the signal can lead to observable BSM effects in neutrino sector, which inadvertently shows up in the DM experiments. 

\section{Summary and Conclusion}
\label{Summary and Conclusion}

 In this article, we have studied the new physics contribution from leptophilic $U(1)_{e-\mu}$ ,$U(1)_{e-\tau}$  and $U(1)_{\mu-\tau}$ models to the CE$\nu$NS, eventually leading to an enhancement to the neutrino floor, which is soon going to become sensitive to the future DM direct detection experiments. 
We have included the latest combined constraints from electron beam dump experiments, neutrino scattering experiments and astrophysical constraints etc., on these models, to find out relatively relaxed constraints on the $U(1)_{\mu-\tau}$ model. The enhancement in the CE$\nu$NS process for these models are confronted with combined experimental constraints on the $m_{Z'} -g_{i-j}$ plane. We were able to achieve 50\% and more enhancement in CE$\nu$NS for the case of $U(1)_{\mu-\tau}$ model compared to that of the SM, in the allowed parameter space with $Z'$ mass hovering in the range of 10-50 MeV.  Due to tighter constraints from $\bar{\nu}-e^-$ scattering cross-section measured by TEXONO  for the $U(1)_{e-\mu}$ and $U(1)_{e-\tau}$ models in the same $m_{Z'} -g_{i-j}$ parameter space , we could not manage any sizable ($\geq 5\% $) enhancement. 
For $m_{Z'}, g_{\mu-\tau}$ values showing the maximum augmentation in the allowed region for $U(1)_{\mu-\tau}$ model we pick a benchmark point $m_{Z'} = 19 \rm MeV, g_{\mu-\tau} = 8 \times 10^{-4} $. We have shown contours of neutrino-nucleus scattering event rate with its variation with nuclear recoil to pin down the rate enhancement compared against SM. For that benchmark point, neutrino-nucleus rate amplification by factors of 2.8  and 1.8 were seen for the cases of Germanium and Xenon respectively, at nuclear recoil energies around 0.01 keV, which diminishes at higher recoil energies. This enhancement is a combination of increase of neutrino-nucleus scattering rate for $U(1)_{\mu-\tau}$, further weighted by the neutrino flux. 
Finally for the neutrino floor, first the exposure required to produce one neutrino-nucleus scattering events for a given threshold energy in the DM direct detection detectors, is obtained. The same exposure is used to investigate the contribution of $U(1)_{\mu-\tau}$ model in the contours depicting values of DM-nucleon scattering cross-section ($\sigma^0_n$) for DM masses ($m_{DM}$) above which we can be certain at 90\% confidence level (2.3 DM events per 1 neutrino scattering event), that measured events are coming from the DM scattering with detecting material rather than neutrino scattering. Enhancement by a factor 2.8 and 1.82 in the neutrino floor were respectively seen for Germanium and Xenon based experiments, in the lighter DM region with $m_{\rm DM} < 10~$GeV. 

In conclusion we find that $U(1)_{\mu-\tau}$ provides significant modification in the CE$\nu$NS floor. This enhancement is especially significant for low mass (less than 10 GeV) dark matter. From the context of DM extension to $U(1)_{\mu-\tau}$ model, the enhancement is noteworthy as the parameter space which leads to the maximum enhancement, can also explain anomalous magnetic moment of muon and relic density of dark matter simultaneously. Therefore, it can be worthwhile to probe the parameter region in neutrino scattering experiments, like COHERENT experiment, to get  a clear picture of the impact the model has in BSM neutrino physics. Further, DM direct detection experiments can reach the enhanced neutrino floor according to the future projections which may ultimately enable us to probe the hitherto unknown neutrino flux in the DM experiments.

\section*{Acknowledgments}
 We would like to thank Prof. Debajyoti Choudhury and Prof. Subhendra Mohanty for useful discussions.
SS thanks UGC for the DS Kothari postdoctoral fellowship grant with award letter No.F.4-2/2006 (BSR)/PH/17-18/0126. MPS would like to thank Prof. Amitabh Mukherjee and Prof. T.R. Seshadri for their support. MPS thanks CSIR JRF fellowship and SERB core research grant CRG/ 2018/ 004889 for partial financial support.

\appendix

\section*{Appendix}

\subsection{Enhancement in CE$\nu$NS event rate and neutrino floor: few benchmark scenarios }
\begin{table}[h]
\centering
\begin{tabular}{ |c|c|c|c|c| } 
\hline
E$_r$(keV) & $\frac{d R^{SM}_{\nu-N}}{d E_r}$(Ton Year keV)$^{-1}$ &  $\frac{d R^{SM}_{\nu-N}}{d E_r}$ (Ton Year keV)$^{-1}$. & $\frac{\frac{d R^{SM}_{\nu-N}}{d E_r}} { \frac{d R^{SM}_{\nu-N}}{d E_r}}$ \\
\hline
0.001 & 1.95644$\times 10^6$ & 5.1514$  \times 10^6$ & 2.63  \\
\hline 
0.005 & 2.76$\times 10^5$ & 7.34$\times 10^5$ &2.65 \\ 
\hline
0.01 & 2.01$\times 10^5$ & 5.34$\times 10^5$ & 2.65 \\ 
\hline
0.05 & 3918.8 & 10525.8 & 2.68\\
\hline 
0.1 & 258.2 & 800.8 & 3.10\\ 
\hline
0.5 & 151.8 & 428.7 & 2.82\\ 
\hline
1.0 & 83.3 & 212.3 & 2.54\\
\hline 
5.0 & 0.18 & 0.31 &1.67 \\ 
\hline
10.0 & 2.9$\times 10^{-4}$ & 3.9$\times 10^{-4}$ & 1.34 \\
\hline
\end{tabular}
\caption{ \small \em{Neutrino nucleus event rate versus recoil energy table showing the comparison between U(1)$_{\mu-\tau}$  and SM   for Germanium nuclei.  Benchmark
chosen: $Z'$  mass 19 MeV and coupling  $g_{\mu-\tau}= 8.0\times10^{-4}$.}}
\label{table:1}
\end{table}
\begin{table}[h]
\centering
\begin{tabular}{ |c|c|c|c|c|} 
\hline
E$_r$(keV) & $\frac{d R^{SM}_{\nu-N}}{d E_r}$(Ton Year keV)$^{-1}$ & $\frac{d R^{\mu-\tau}_{\nu-N}}{d E_r}$ (Ton Year keV)$^{-1}$.  & $\frac{\frac{d R^{SM}_{\nu-N}}{d E_r}} { \frac{d R^{SM}_{\nu-N}}{d E_r}}$ \\
\hline
0.001 & 6.746$\times 10^6$ & 1.241$\times 10^7$ & 1.84 \\
\hline 
0.005 & 1.866$\times 10^6$& 2.377$\times 10^6$ & 1.84\\ 
\hline
0.01 & 4.48$\times 10^5$ & 8.28$\times 10^6$ & 1.84 \\ 
\hline
0.05 & 1613 & 3360 & 2.08\\
\hline 
0.1 & 1401 & 2837 & 2.04 \\ 
\hline
0.5 & 528 & 965 & 1.82\\ 
\hline
1.0 & 159 & 263.3 & 1.64 \\
\hline 
5.0 & 0.0016 & 0.0019 & 1.20\\ 
\hline
10.0 & 5.43$\times 10^{-4}$ & 5.99$\times 10^{-4}$ & 1.10\\ 
\hline
\end{tabular}
\caption{\small \em{ Neutrino nucleus event rate versus recoil energy table showing the comparison between U(1)$_{\mu-\tau}$ model and SM   for Xenon nuclei.  Benchmark
chosen: $Z'$  mass 19 MeV and coupling  $g_{\mu-\tau}= 8.0\times10^{-4}$.}}
\label{table:2}
\end{table}
\newpage
\begin{table}[h]
\centering
\begin{tabular}{ |c|c|c|c|c| } 
\hline
m$_{DM}$(GeV) & SM Neutrino floor(cm$^{2}$) &  U(1)$_{\mu-\tau}$ Neutrino floor(cm$^{2}$) & Enhancement \\
\hline
0.5 & 2.37$\times 10^{-43}$ & 6.30$\times 10^{-43}$  & 2.65  \\
\hline 
1. &  5.30$\times 10^{-44}$ & 1.42$\times 10^{-43}$  & 2.67  \\ 
\hline
5 & 7.11$\times 10^{-45}$ & 1.90$\times 10^{-44}$  & 2.67  \\ 
\hline
10 & 2.16$\times 10^{-47}$ & 2.87$\times 10^{-47}$  & 1.32 \\
\hline 
50 & 3.90$\times 10^{-49}$ & 4.36$\times 10^{-49}$  & 1.11 \\ 
\hline
100 & 4.08$\times 10^{-49}$ & 4.53$\times 10^{-49}$  & 1.11 \\ 
\hline
500 & 1.23$\times 10^{-48}$ & 1.33$\times 10^{-48}$  & 1.08 \\
\hline 
1000 & 2.31$\times 10^{-48}$ & 2.48$\times 10^{-48}$  & 1.07 \\ 
\hline
\end{tabular}
\caption{ \small \em{Neutrino floor versus dark matter mass table highlighting modification of neutrino floor for U(1)$_{\mu-\tau}$  with respect to SM  for Germanium nuclei. Benchmark
chosen: $Z'$  mass 19 MeV and coupling  $g_{\mu-\tau}= 8.0\times10^{-4}$.} }
\label{table:3}
\end{table}

\begin{table}[h]
\centering
\begin{tabular}{ |c|c|c|c|c| } 
\hline
m$_{DM}$(GeV) & SM Neutrino floor(cm$^{2}$) &  U(1)$_{\mu-\tau}$ Neutrino floor(cm$^{2}$) & Enhancement \\
\hline
0.5 & 3.92$\times 10^{-43}$ & 7.27$\times 10^{-43}$  & 1.85  \\
\hline 
1. &  8.67$\times 10^{-44}$ & 1.62$\times 10^{-43}$  & 1.86  \\ 
\hline
5 & 1.10$\times 10^{-44}$ & 2.06$\times 10^{-44}$  & 1.87  \\ 
\hline
10 & 1.51$\times 10^{-47}$ & 1.78$\times 10^{-47}$  & 1.17 \\
\hline 
50 & 3.41$\times 10^{-49}$ & 3.66$\times 10^{-49}$  & 1.11 \\ 
\hline
100 & 4.03$\times 10^{-49}$ & 4.25$\times 10^{-49}$  & 1.07 \\ 
\hline
500 & 6.83$\times 10^{-48}$ & 6.89$\times 10^{-48}$  & 1.01 \\
\hline 
1000 & 1.04$\times 10^{-48}$ & 1.05$\times 10^{-48}$  & 1.01 \\ 
\hline
\end{tabular}
\caption{ \small \em{Neutrino floor versus dark matter mass table highlighting modification of neutrino floor for U$_{\mu-\tau}$  with respect to SM  for Xenon nuclei. Benchmark
chosen: $Z'$  mass 19 MeV and coupling  $g_{\mu-\tau}= 8.0\times10^{-4}$.} }\label{table:4}
\end{table}

\subsection{$Z'-\gamma$ mixing in U$_{i-j}$ model}

\begin{figure}[h!]
\centering 
\begin{subfigure}{.4\textwidth}\centering
  \includegraphics[width=\columnwidth]{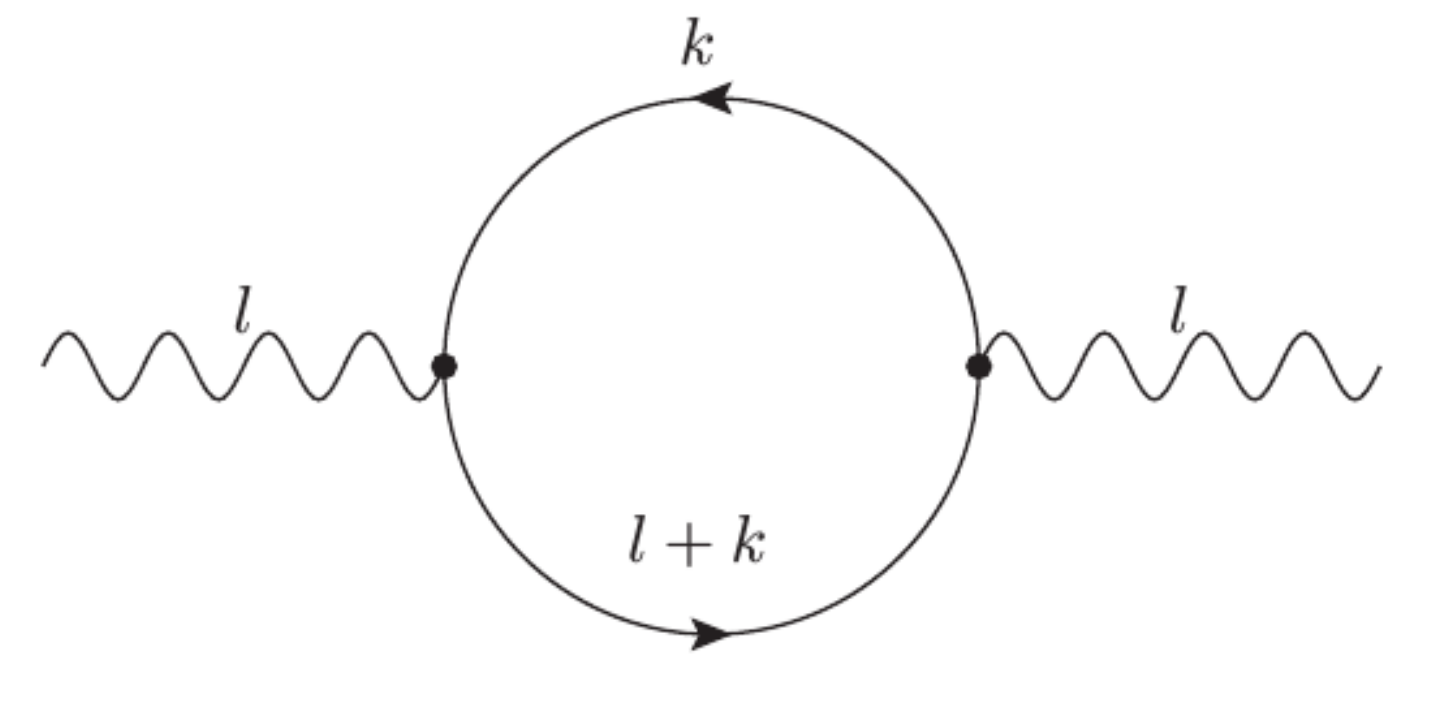}
\end{subfigure}%
\caption{Lepton loop through which $Z^{\prime}-Z/\gamma$ mixing is induced in U$_{i-j}$ model.} 
\label{feynloop}
\end{figure}

In the $U(1)_{i-j}$ models, contribution to CE$\nu$NS due to extra $Z^{\prime}$ boson are mediated by $Z^{\prime}-Z/\gamma$ mixing as shown in Fig.~\ref{feyn2}. The loop (see Fig. \ref{feynloop}) contribution driven mixing element is given by, 
\begin{eqnarray}
&&\mu^{2\epsilon} \int \frac{d^d k}{(2\pi)^d}\frac{2tr\left[\gamma^\mu(\slashed{l+k}+m_l)\gamma^\nu(\slashed{k}+m_l)\right]}{\left[(l+k)^2-m^2_l\right]\left[k^2-m^2_l\right]}\label{loopfactor}
\end{eqnarray}
Numerator can be simplified as,
\begin{eqnarray}
 && tr\left[\gamma^\mu(\slashed{l+k}+m_l)\gamma^\nu(\slashed{k}+m_l)\right]\nonumber
\\\nonumber
&&= tr\left[\gamma^\mu \slashed{l} \gamma^\nu \slashed{k}+\gamma^\mu \slashed{k}\gamma^\nu \slashed{k}+ m^2_l \gamma^\mu \gamma^\nu\right]
\\
&&=4\left[l^\mu k^\nu + l^\nu k^\mu - g^{\mu \nu}(l.k) + 2 k^\mu k^\nu - g^{\mu \nu}(k^2) + m^2_l g^{\mu\nu}\right]\label{numerator}
\end{eqnarray}
{Using Feynman parametrization}
\begin{eqnarray}
&&\frac{1}{\left[(l+k)^2-m^2_l\right]\left[k^2-m^2_l\right]}
=\int dx \frac{1}{\big(x\left((l+k)^2-m^2_l\right)-(1-x)\left(k^2-m^2_l\right)\big)^2}\label{feynpara}
\end{eqnarray}
Putting back \ref{feynpara} and \ref{numerator} in \ref{loopfactor}  
\begin{eqnarray}
&&\mu^{2\epsilon} \int_0 ^1  dx \int \frac{d^d k}{(2\pi)^d}\frac{4\left[l^\mu k^\nu + l^\nu k^\mu - g^{\mu \nu}(l.k) + 2 k^\mu k^\nu - g^{\mu \nu}(k^2) + m^2_l g^{\mu\nu}\right]}{\big(x\left((l+k)^2-m^2_l\right)-(1-x)\left(k^2-m^2_l\right)\big)^2}
\\\nonumber
\end{eqnarray}

shifting  $k \rightarrow k'-lx$  and substituting $\Delta= x(x-1)l^2 +m_l^2$,We have

\begin{eqnarray}
&&\mu^{2\epsilon} \int_0 ^1  dx \int \frac{d^d k}{(2\pi)^d}\frac{4[(x - x^2)(-2 l^\nu l^\mu + g^{\mu\nu} l^2)  +(\frac{2}{d}-1)g^{\mu\nu}k'^2 + m^2_l g^{\mu\nu} ]}{(k'^2-\Delta)^2}
\end{eqnarray}
Under the rotation $k^0\rightarrow i k_{E}$ and $k^i\rightarrow  k^i_{E}$
\begin{eqnarray}
&&i \mu^{2\epsilon}\int^1_0 dx \int \frac{d^d k_E}{(2\pi)^d}\frac{4[(x - x^2)(-2 l^\nu l^\mu + g^{\mu\nu} l^2) -(\frac{2}{d}-1)g^{\mu\nu}k_E^2 + m^2_l g^{\mu\nu} ]}{(k_E^2+\Delta)^2}\label{loop1}
\end{eqnarray}
using the simplification
\begin{eqnarray}
&& \frac{d^d k_E}{(2\pi)^d}= \frac{k^{d-1}_E d k_E}{(2\pi)^d}d\Omega_d\nonumber
\\
&& \int_0^\infty dy \frac{y^a}{(y^2 + \Delta)^b}= \Delta^{\frac{a+1}{2}-b}\frac{\Gamma(\frac{a+1}{2})\Gamma(b-\frac{a+1}{2})}{2\Gamma(b)}\nonumber,
\end{eqnarray}

where, $\int\Omega_d= \frac{2\pi^{\frac{d}{2}}}{\Gamma (d/2)}$ 
\\
\\
Integral in \ref{loop1} in  $\overline{MS}$ scheme is solved as.
\begin{eqnarray}
&&=\frac{1}{(2\pi^2)}[- l^\mu l^\nu + g^{\mu\nu} l^2]\int^1_0 dx\,(-\frac{2}{\epsilon} +\log\frac{x(x-1)l^2 + m^2_l}{4\pi\mu^4}+\gamma_E)x(1-x)\nonumber
\end{eqnarray}
In U(1)$_{i -j}$ model the infinite terms cancel between two lepton flavors and we have
\\
\begin{eqnarray}
&&\frac{1}{(2\pi^2)}[- l^\mu l^\nu + g^{\mu\nu} l^2]\int^1_0 dx\,( \log\frac{x(x-1)l^2 + m^2_{l_i}}{x(x-1)l^2 + m^2_{l_j}})x(1-x)
\end{eqnarray}

\bibliographystyle{unsrtnat}
\bibliography{ref}

\begin{thebibliography}{49}
\providecommand{\natexlab}[1]{#1}
\providecommand{\url}[1]{\texttt{#1}}
\expandafter\ifx\csname urlstyle\endcsname\relax
  \providecommand{\doi}[1]{doi: #1}\else
  \providecommand{\doi}{doi: \begingroup \urlstyle{rm}\Url}\fi

\bibitem[Aprile et~al.(2016)]{Aprile:2015uzo}
E.~Aprile et~al.
\newblock {Physics reach of the XENON1T dark matter experiment}.
\newblock \emph{JCAP}, 1604\penalty0 (04):\penalty0 027, 2016.
\newblock \doi{10.1088/1475-7516/2016/04/027}.

\bibitem[Cui et~al.(2017)]{Cui:2017nnn}
Xiangyi Cui et~al.
\newblock {Dark Matter Results From 54-Ton-Day Exposure of PandaX-II
  Experiment}.
\newblock \emph{Phys. Rev. Lett.}, 119\penalty0 (18):\penalty0 181302, 2017.
\newblock \doi{10.1103/PhysRevLett.119.181302}.

\bibitem[Akerib et~al.(2017)]{Akerib:2016vxi}
D.S. Akerib et~al.
\newblock {Results from a search for dark matter in the complete LUX exposure}.
\newblock \emph{Phys. Rev. Lett.}, 118\penalty0 (2):\penalty0 021303, 2017.
\newblock \doi{10.1103/PhysRevLett.118.021303}.

\bibitem[Agnese et~al.(2013)]{Agnese:2013rvf}
R.~Agnese et~al.
\newblock {Silicon Detector Dark Matter Results from the Final Exposure of CDMS
  II}.
\newblock \emph{Phys. Rev. Lett.}, 111\penalty0 (25):\penalty0 251301, 2013.
\newblock \doi{10.1103/PhysRevLett.111.251301}.

\bibitem[Akimov et~al.(2017)]{Akimov:2017ade}
D.~Akimov et~al.
\newblock {Observation of Coherent Elastic Neutrino-Nucleus Scattering}.
\newblock \emph{Science}, 357\penalty0 (6356):\penalty0 1123--1126, 2017.
\newblock \doi{10.1126/science.aao0990}.

\bibitem[Bauer et~al.(2018)Bauer, Foldenauer, and Jaeckel]{Bauer:2018onh}
Martin Bauer, Patrick Foldenauer, and Joerg Jaeckel.
\newblock {Hunting All the Hidden Photons}.
\newblock \emph{JHEP}, 07:\penalty0 094, 2018.
\newblock \doi{10.1007/JHEP07(2018)094}.
\newblock [JHEP18,094(2020)].

\bibitem[Chang et~al.(2017)]{TheDAMPE:2017dtc}
J.~Chang et~al.
\newblock {The DArk Matter Particle Explorer mission}.
\newblock \emph{Astropart. Phys.}, 95:\penalty0 6--24, 2017.
\newblock \doi{10.1016/j.astropartphys.2017.08.005}.

\bibitem[Aguilar(2013)]{PhysRevLett.110.141102}
M.~.et~all Aguilar.
\newblock First result from the alpha magnetic spectrometer on the
  international space station: Precision measurement of the positron fraction
  in primary cosmic rays of 0.5--350 gev.
\newblock \emph{Phys. Rev. Lett.}, 110:\penalty0 141102, Apr 2013.
\newblock \doi{10.1103/PhysRevLett.110.141102}.
\newblock URL \url{https://link.aps.org/doi/10.1103/PhysRevLett.110.141102}.

\bibitem[Chao et~al.(2019)Chao, Jiang, Wang, and Zhang]{Chao:2019pyh}
Wei Chao, Jian-Guo Jiang, Xuan Wang, and Xing-Yu Zhang.
\newblock {Direct Detections of Dark Matter in the Presence of Non-standard
  Neutrino Interactions}.
\newblock \emph{JCAP}, 1908:\penalty0 010, 2019.
\newblock \doi{10.1088/1475-7516/2019/08/010}.

\bibitem[Heeck et~al.(2019)Heeck, Lindner, Rodejohann, and Vogl]{Heeck:2018nzc}
Julian Heeck, Manfred Lindner, Werner Rodejohann, and Stefan Vogl.
\newblock {Non-Standard Neutrino Interactions and Neutral Gauge Bosons}.
\newblock \emph{SciPost Phys.}, 6\penalty0 (3):\penalty0 038, 2019.
\newblock \doi{10.21468/SciPostPhys.6.3.038}.

\bibitem[Bœhm et~al.(2019)Bœhm, Cerdeño, Machado, Olivares-Del~Campo,
  Perdomo, and Reid]{Boehm:2018sux}
C.~Bœhm, D.~G. Cerdeño, P.~A.~N. Machado, A.~Olivares-Del~Campo, E.~Perdomo,
  and E.~Reid.
\newblock {How high is the neutrino floor?}
\newblock \emph{JCAP}, 1901\penalty0 (01):\penalty0 043, 2019.
\newblock \doi{10.1088/1475-7516/2019/01/043}.

\bibitem[Bertuzzo et~al.(2017)Bertuzzo, Deppisch, Kulkarni, Perez~Gonzalez, and
  Zukanovich~Funchal]{Bertuzzo:2017tuf}
Enrico Bertuzzo, Frank~F. Deppisch, Suchita Kulkarni, Yuber~F. Perez~Gonzalez,
  and Renata Zukanovich~Funchal.
\newblock {Dark Matter and Exotic Neutrino Interactions in Direct Detection
  Searches}.
\newblock 1 2017.
\newblock \doi{10.1007/JHEP04(2017)073}.
\newblock [Erratum: JHEP 04, 073 (2017)].

\bibitem[Foot(1991)]{Foot:1990mn}
Robert Foot.
\newblock {New Physics From Electric Charge Quantization?}
\newblock \emph{Mod. Phys. Lett.}, A6:\penalty0 527--530, 1991.
\newblock \doi{10.1142/S0217732391000543}.

\bibitem[He et~al.(1991{\natexlab{a}})He, Joshi, Lew, and Volkas]{He:1990pn}
X.~G. He, Girish~C. Joshi, H.~Lew, and R.~R. Volkas.
\newblock {NEW Z-prime PHENOMENOLOGY}.
\newblock \emph{Phys. Rev.}, D43:\penalty0 22--24, 1991{\natexlab{a}}.
\newblock \doi{10.1103/PhysRevD.43.R22}.

\bibitem[He et~al.(1991{\natexlab{b}})He, Joshi, Lew, and Volkas]{He:1991qd}
Xiao-Gang He, Girish~C. Joshi, H.~Lew, and R.~R. Volkas.
\newblock {Simplest Z-prime model}.
\newblock \emph{Phys. Rev.}, D44:\penalty0 2118--2132, 1991{\natexlab{b}}.
\newblock \doi{10.1103/PhysRevD.44.2118}.

\bibitem[Choudhury et~al.(2020)Choudhury, Deka, Mandal, and
  Sadhukhan]{Choudhury:2020cpm}
Debajyoti Choudhury, Kuldeep Deka, Tanumoy Mandal, and Soumya Sadhukhan.
\newblock {Neutrino and $Z'$ phenomenology in an anomaly-free $\mathbf{U}(1)$
  extension: role of higher-dimensional operators}.
\newblock \emph{JHEP}, 06:\penalty0 111, 2020.
\newblock \doi{10.1007/JHEP06(2020)111}.

\bibitem[Bi et~al.(2009)Bi, He, and Yuan]{Bi:2009uj}
Xiao-Jun Bi, Xiao-Gang He, and Qiang Yuan.
\newblock {Parameters in a class of leptophilic models from PAMELA, ATIC and
  FERMI}.
\newblock \emph{Phys. Lett.}, B678:\penalty0 168--173, 2009.
\newblock \doi{10.1016/j.physletb.2009.06.009}.

\bibitem[Arcadi et~al.(2018)Arcadi, Hugle, and Queiroz]{Arcadi:2018tly}
Giorgio Arcadi, Thomas Hugle, and Farinaldo~S. Queiroz.
\newblock {The Dark $L_\mu - L_\tau$ Rises via Kinetic Mixing}.
\newblock \emph{Phys. Lett.}, B784:\penalty0 151--158, 2018.
\newblock \doi{10.1016/j.physletb.2018.07.028}.

\bibitem[Foldenauer(2019)]{Foldenauer:2018zrz}
Patrick Foldenauer.
\newblock {Light dark matter in a gauged $U(1)_{L_\mu-L_\tau}$ model}.
\newblock \emph{Phys. Rev.}, D99\penalty0 (3):\penalty0 035007, 2019.
\newblock \doi{10.1103/PhysRevD.99.035007}.

\bibitem[Altmannshofer et~al.(2016)Altmannshofer, Gori, Profumo, and
  Queiroz]{Altmannshofer:2016jzy}
Wolfgang Altmannshofer, Stefania Gori, Stefano Profumo, and Farinaldo~S.
  Queiroz.
\newblock {Explaining dark matter and B decay anomalies with an $L_\mu -
  L_\tau$ model}.
\newblock \emph{JHEP}, 12:\penalty0 106, 2016.
\newblock \doi{10.1007/JHEP12(2016)106}.

\bibitem[Patra et~al.(2017)Patra, Rao, Sahoo, and Sahu]{Patra:2016shz}
Sudhanwa Patra, Soumya Rao, Nirakar Sahoo, and Narendra Sahu.
\newblock {Gauged $U(1)_{L_\mu - L_\tau}$ model in light of muon $g-2$ anomaly,
  neutrino mass and dark matter phenomenology}.
\newblock \emph{Nucl. Phys.}, B917:\penalty0 317--336, 2017.
\newblock \doi{10.1016/j.nuclphysb.2017.02.010}.

\bibitem[Baek(2016)]{Baek:2015fea}
Seungwon Baek.
\newblock {Dark matter and muon $(g-2)$ in local $U(1)_{L_\mu-L_\tau}$-extended
  Ma Model}.
\newblock \emph{Phys. Lett.}, B756:\penalty0 1--5, 2016.
\newblock \doi{10.1016/j.physletb.2016.02.062}.

\bibitem[Duan et~al.(2018)Duan, He, Wu, and Yang]{Duan:2017qwj}
Guang~Hua Duan, Xiao-Gang He, Lei Wu, and Jin~Min Yang.
\newblock {Leptophilic dark matter in gauged $U(1)_{L{_e}-L_{\mu }}$ model in
  light of DAMPE cosmic ray ${e{^+}} + {e{^-}}$ excess}.
\newblock \emph{Eur. Phys. J.}, C78\penalty0 (4):\penalty0 323, 2018.
\newblock \doi{10.1140/epjc/s10052-018-5805-1}.

\bibitem[Baek(2019)]{Baek:2019qte}
Seungwon Baek.
\newblock {Scalar dark matter behind $b \to s \mu \mu$ anomaly}.
\newblock \emph{JHEP}, 05:\penalty0 104, 2019.
\newblock \doi{10.1007/JHEP05(2019)104}.

\bibitem[Riordan et~al.(1987)]{Riordan:1987aw}
E.~M. Riordan et~al.
\newblock {A Search for Short Lived Axions in an Electron Beam Dump
  Experiment}.
\newblock \emph{Phys. Rev. Lett.}, 59:\penalty0 755, 1987.
\newblock \doi{10.1103/PhysRevLett.59.755}.

\bibitem[Bjorken et~al.(1988)Bjorken, Ecklund, Nelson, Abashian, Church, Lu,
  Mo, Nunamaker, and Rassmann]{Bjorken:1988as}
J.~D. Bjorken, S.~Ecklund, W.~R. Nelson, A.~Abashian, C.~Church, B.~Lu, L.~W.
  Mo, T.~A. Nunamaker, and P.~Rassmann.
\newblock {Search for Neutral Metastable Penetrating Particles Produced in the
  SLAC Beam Dump}.
\newblock \emph{Phys. Rev.}, D38:\penalty0 3375, 1988.
\newblock \doi{10.1103/PhysRevD.38.3375}.

\bibitem[Bross et~al.(1991)Bross, Crisler, Pordes, Volk, Errede, and
  Wrbanek]{Bross:1989mp}
A.~Bross, M.~Crisler, Stephen~H. Pordes, J.~Volk, S.~Errede, and J.~Wrbanek.
\newblock {A Search for Shortlived Particles Produced in an Electron Beam
  Dump}.
\newblock \emph{Phys. Rev. Lett.}, 67:\penalty0 2942--2945, 1991.
\newblock \doi{10.1103/PhysRevLett.67.2942}.

\bibitem[Bellini et~al.(2011)]{Bellini:2011rx}
G.~Bellini et~al.
\newblock {Precision measurement of the 7Be solar neutrino interaction rate in
  Borexino}.
\newblock \emph{Phys. Rev. Lett.}, 107:\penalty0 141302, 2011.
\newblock \doi{10.1103/PhysRevLett.107.141302}.

\bibitem[Deniz et~al.(2010)]{Deniz:2009mu}
M.~Deniz et~al.
\newblock {Measurement of Nu(e)-bar -Electron Scattering Cross-Section with a
  CsI(Tl) Scintillating Crystal Array at the Kuo-Sheng Nuclear Power Reactor}.
\newblock \emph{Phys. Rev.}, D81:\penalty0 072001, 2010.
\newblock \doi{10.1103/PhysRevD.81.072001}.

\bibitem[Ballett et~al.(2019)Ballett, Hostert, Pascoli, Perez-Gonzalez,
  Tabrizi, and Zukanovich~Funchal]{Ballett:2019xoj}
Peter Ballett, Matheus Hostert, Silvia Pascoli, Yuber~F. Perez-Gonzalez, Zahra
  Tabrizi, and Renata Zukanovich~Funchal.
\newblock {$Z^\prime$s in neutrino scattering at DUNE}.
\newblock \emph{Phys. Rev. D}, 100\penalty0 (5):\penalty0 055012, 2019.
\newblock \doi{10.1103/PhysRevD.100.055012}.

\bibitem[Vilain et~al.(1994)]{Vilain:1994qy}
P.~Vilain et~al.
\newblock {Precision measurement of electroweak parameters from the scattering
  of muon-neutrinos on electrons}.
\newblock \emph{Phys. Lett.}, B335:\penalty0 246--252, 1994.
\newblock \doi{10.1016/0370-2693(94)91421-4}.

\bibitem[Wise and Zhang(2018)]{Wise:2018rnb}
Mark~B. Wise and Yue Zhang.
\newblock {Lepton Flavorful Fifth Force and Depth-dependent Neutrino Matter
  Interactions}.
\newblock \emph{JHEP}, 06:\penalty0 053, 2018.
\newblock \doi{10.1007/JHEP06(2018)053}.

\bibitem[Dror(2020)]{Dror:2020fbh}
Jeff~A. Dror.
\newblock {Discovering leptonic forces using nonconserved currents}.
\newblock \emph{Phys. Rev. D}, 101\penalty0 (9):\penalty0 095013, 2020.
\newblock \doi{10.1103/PhysRevD.101.095013}.

\bibitem[Freedman(1974)]{Freedman:1973yd}
Daniel~Z. Freedman.
\newblock {Coherent Neutrino Nucleus Scattering as a Probe of the Weak Neutral
  Current}.
\newblock \emph{Phys. Rev.}, D9:\penalty0 1389--1392, 1974.
\newblock \doi{10.1103/PhysRevD.9.1389}.

\bibitem[Lewin and Smith(1996)]{Lewin:1995rx}
J.~D. Lewin and P.~F. Smith.
\newblock {Review of mathematics, numerical factors, and corrections for dark
  matter experiments based on elastic nuclear recoil}.
\newblock \emph{Astropart. Phys.}, 6:\penalty0 87--112, 1996.
\newblock \doi{10.1016/S0927-6505(96)00047-3}.

\bibitem[Akimov et~al.(2015)]{Akimov:2015nza}
D.~Akimov et~al.
\newblock {The COHERENT Experiment at the Spallation Neutron Source}.
\newblock 9 2015.

\bibitem[Billard et~al.(2018)Billard, Johnston, and Kavanagh]{Billard:2018jnl}
Julien Billard, Joseph Johnston, and Bradley~J. Kavanagh.
\newblock {Prospects for exploring New Physics in Coherent Elastic
  Neutrino-Nucleus Scattering}.
\newblock \emph{JCAP}, 11:\penalty0 016, 2018.
\newblock \doi{10.1088/1475-7516/2018/11/016}.

\bibitem[Gonzalez-Garcia et~al.(2018)Gonzalez-Garcia, Maltoni, Perez-Gonzalez,
  and Zukanovich~Funchal]{Gonzalez-Garcia:2018dep}
M.C. Gonzalez-Garcia, Michele Maltoni, Yuber~F. Perez-Gonzalez, and Renata
  Zukanovich~Funchal.
\newblock {Neutrino Discovery Limit of Dark Matter Direct Detection Experiments
  in the Presence of Non-Standard Interactions}.
\newblock \emph{JHEP}, 07:\penalty0 019, 2018.
\newblock \doi{10.1007/JHEP07(2018)019}.

\bibitem[Strigari(2009)]{Strigari:2009bq}
Louis~E. Strigari.
\newblock {Neutrino Coherent Scattering Rates at Direct Dark Matter Detectors}.
\newblock \emph{New J. Phys.}, 11:\penalty0 105011, 2009.
\newblock \doi{10.1088/1367-2630/11/10/105011}.

\bibitem[Billard et~al.(2014)Billard, Strigari, and
  Figueroa-Feliciano]{Billard:2013qya}
J.~Billard, L.~Strigari, and E.~Figueroa-Feliciano.
\newblock {Implication of neutrino backgrounds on the reach of next generation
  dark matter direct detection experiments}.
\newblock \emph{Phys. Rev.}, D89\penalty0 (2):\penalty0 023524, 2014.
\newblock \doi{10.1103/PhysRevD.89.023524}.

\bibitem[Hernandez(2010)]{Hernandez:2010mi}
P.~Hernandez.
\newblock {Neutrino physics}.
\newblock In \emph{{High-energy physics. Proceedings, 5th CERN-Latin-American
  School, Recinto Quirama, Colombia, March 15-28, 2009}}, 2010.

\bibitem[Lopes and Turck-Chièze(2013)]{Lopes:2013nfa}
Ilídio Lopes and Sylvaine Turck-Chièze.
\newblock {Solar neutrino physics oscillations: Sensitivity to the electronic
  density in the Sun's core}.
\newblock \emph{Astrophys. J.}, 765:\penalty0 14, 2013.
\newblock \doi{10.1088/0004-637X/765/1/14}.

\bibitem[Read(2014)]{Read:2014qva}
J.~I. Read.
\newblock {The Local Dark Matter Density}.
\newblock \emph{J. Phys.}, G41:\penalty0 063101, 2014.
\newblock \doi{10.1088/0954-3899/41/6/063101}.

\bibitem[Pato et~al.(2015)Pato, Iocco, and Bertone]{Pato:2015dua}
Miguel Pato, Fabio Iocco, and Gianfranco Bertone.
\newblock {Dynamical constraints on the dark matter distribution in the Milky
  Way}.
\newblock \emph{JCAP}, 1512\penalty0 (12):\penalty0 001, 2015.
\newblock \doi{10.1088/1475-7516/2015/12/001}.

\bibitem[de~Salas(2019)]{deSalas:2019rdi}
Pablo~F. de~Salas.
\newblock {Dark matter local density determination based on recent
  observations}.
\newblock In \emph{{16th International Conference on Topics in Astroparticle
  and Underground Physics (TAUP 2019) Toyama, Japan, September 9-13, 2019}},
  2019.

\bibitem[Wyenberg and Shoemaker(2018)]{Wyenberg:2018eyv}
Jason Wyenberg and Ian~M. Shoemaker.
\newblock {Mapping the neutrino floor for direct detection experiments based on
  dark matter-electron scattering}.
\newblock \emph{Phys. Rev. D}, 97\penalty0 (11):\penalty0 115026, 2018.
\newblock \doi{10.1103/PhysRevD.97.115026}.

\bibitem[Marrodán~Undagoitia and Rauch(2016)]{Undagoitia:2015gya}
Teresa Marrodán~Undagoitia and Ludwig Rauch.
\newblock {Dark matter direct-detection experiments}.
\newblock \emph{J. Phys.}, G43\penalty0 (1):\penalty0 013001, 2016.
\newblock \doi{10.1088/0954-3899/43/1/013001}.

\bibitem[Agnese et~al.(2017)]{Agnese:2016cpb}
R.~Agnese et~al.
\newblock {Projected Sensitivity of the SuperCDMS SNOLAB experiment}.
\newblock \emph{Phys. Rev.}, D95\penalty0 (8):\penalty0 082002, 2017.
\newblock \doi{10.1103/PhysRevD.95.082002}.

\bibitem[Agnese et~al.(2018)]{Agnese:2017jvy}
R.~Agnese et~al.
\newblock {Low-mass dark matter search with CDMSlite}.
\newblock \emph{Phys. Rev.}, D97\penalty0 (2):\penalty0 022002, 2018.
\newblock \doi{10.1103/PhysRevD.97.022002}.

\end{thebibliography}

\end{document}